\documentclass[aip,jcp,reprint]{revtex4-2}
\usepackage{graphicx}
\usepackage{dcolumn}
\usepackage{bm}
\usepackage{mathrsfs}
\usepackage{amssymb}
\usepackage{amsmath}
\usepackage{stackengine}
\usepackage{color}
\newcommand{\Liou}{\mathcal{L_{}}}

\newcommand{\mathleft}{\@fleqntrue\@mathmargin0pt}
\newcommand{\mathcenter}{\@fleqnfalse}

\allowdisplaybreaks

\DeclareMathOperator{\Tr} { Tr}

\begin{document}

\title{Microscopic theory of a Janus motor in a non-equilibrium fluid: Surface hydrodynamics and boundary conditions}

\author{Bryan Robertson}
%\altaffiliation[Also at ]{Physics Department, XYZ University.}%Lines break automatically or can be forced with \\

\author{Jeremy Schofield}%
\email{jeremy.schofield@utoronto.ca}
\author{Raymond Kapral}%
\email{r.kapral@utoronto.ca}
\affiliation{Chemical Physics Theory Group, Department of Chemistry, University of Toronto, Toronto, Ontario, Canada, M5S 3H6}

\date{\today}

\begin{abstract}
We present a derivation from first principles of the coupled equations of motion of an active self-diffusiophoretic Janus motor and the hydrodynamic densities of its fluid environment that are nonlinearly displaced from equilibrium. The derivation makes use of time-dependent projection operator techniques defined in terms of slowly varying coarse-grained microscopic densities of the fluid species number, total momentum, and energy. The exact equations of motion are simplified using time-scale arguments, resulting in Markovian equations for the Janus motor linear and angular velocities with average forces and torques that depend on the fluid densities. For a large colloid, the fluid equations are separated into bulk and interfacial contributions, and the conditions under which the dynamics of the fluid densities can be accurately represented by bulk hydrodynamic equations subject to boundary conditions on the colloid are determined. We show how the results for boundary conditions based on continuum theory can be obtained from the molecular description and provide Green-Kubo expressions for all transport coefficients, including the diffusiophoretic coupling and the slip coefficient.
 
\end{abstract}

\maketitle

\section{Introduction} \label{sec:intro}
Self-propelled small synthetic particles have been extensively studied through experiment, theory, and simulation.~\cite{W13,K13,BDLRVV16,IGS17,S19,GK19,G_20} Much of the stimulus for this research activity stems from the potential and actual applications of these active agents, often related to their uses for cargo transport and as vehicles for targeted drug delivery, but there are many other uses.~\cite{xu2019self,Pumera2015,SSK14,APTW14} Such active particles often have micrometer or sub-micrometer dimensions, which is an interesting and theoretically challenging regime lying between the fully microscopic and macroscopic domains.

While active particles can have many different shapes and use different mechanisms for propulsion, a great deal of work has focused on micrometer-sized active Janus colloids propelled by phoretic mechanisms. Here, we are interested in diffusiophoresis where chemical species concentration gradients, either externally imposed or self-generated, are an essential part of the mechanism. Continuum descriptions of colloidal motion by this mechanism are well known and are based on solutions of the Stokes equation coupled to reaction-diffusion equations.~\cite{ALP82,A89,GLA07,OPD17} Continuum models are often appropriate, even for micrometer-sized colloids, although in this domain fluctuations are important and may be treated using fluctuating hydrodynamic methods.~\cite{GK18a} In both cases, an essential part of the theory is the application of suitable boundary conditions for the fluid fields on the colloid surface. As one proceeds to the sub-micrometer regime, especially for particles with order nanometer sizes, the continuum description will lose its validity and microscopic descriptions must be used. Active colloids in this nanometer regime have been studied experimentally and through simulation.~\cite{LAMHGF14,CK14,Wilson2016}

The development of a microscopic description of small diffusiophoretically-active colloids that allows one to pass from the microscopic to macroscopic domains would provide a better understanding of how active motion occurs across these space scales. The results presented in this paper are twofold: First, from a molecular perspective, we construct evolution equations for a Janus colloid and fluid field equations that account for the presence of the active colloid. Second, for a large Janus particle, we derive boundary conditions from the microscopic theory and, in the process, obtain correlation function expressions for the surface transport properties that appear as parameters in phenomenological continuum descriptions. Since active motion occurs only if the system is driven out of equilibrium, we employ a statistical mechanical framework that accounts for non-equilibrium constraints on the system.~\cite{R67,P68,OL79} The resulting generalized hydrodynamic equations may then be simplified in specific time and space regimes. If the colloid is massive compared to the solvent, Brownian motion scaling can be applied to simplify the evolution equations for the active colloid. The equations for the fluid-conserved fields adopt a Markovian form on time scales that are long compared to microscopic times. If the characteristic size of the colloid is large compared to the boundary zone where fluid colloid interactions occur, the colloid's coupling to the fluid can be described through boundary conditions.

The paper begins with a description of the microscopic model of the system, the definition of the colloid and coarse-grained fluid fields whose equations of motion are of interest, and the non-equilibrium ensemble and constraints imposed on these fields. After the colloid and generalized hydrodynamic fluid equations are derived, space and time scale considerations are used to simplify the equations into tractable form. The resulting equations are decomposed into bulk and surface contributions for a large colloid, and boundary conditions are derived from the surface equations. Noteworthy results in this part of the paper are microscopic correlation function expressions for the transport properties that enter the boundary conditions. The discussion and conclusions establish links between this microscopic theory and fluctuating hydrodynamic and continuum methods. Technical details are given in the Appendices.

\section{Microscopic model for a fluid with an active Janus colloid}
\label{sec:system}

We consider an active Janus colloid in a fluid environment at temperature $T$ comprising $N_S$ solvent molecules $S$ and $N_R$ dilutely dispersed reactive molecules $R$ so that $N_S \gg N_R$ and the total number of fluid molecules is $N=N_S+N_R$. All of these species are taken to have mass $m$. The phase space coordinates of the fluid particles are denoted by $\bm{x}$.
\begin{figure}[htbp]
\centering
\resizebox{0.9\columnwidth}{!}{
      \includegraphics{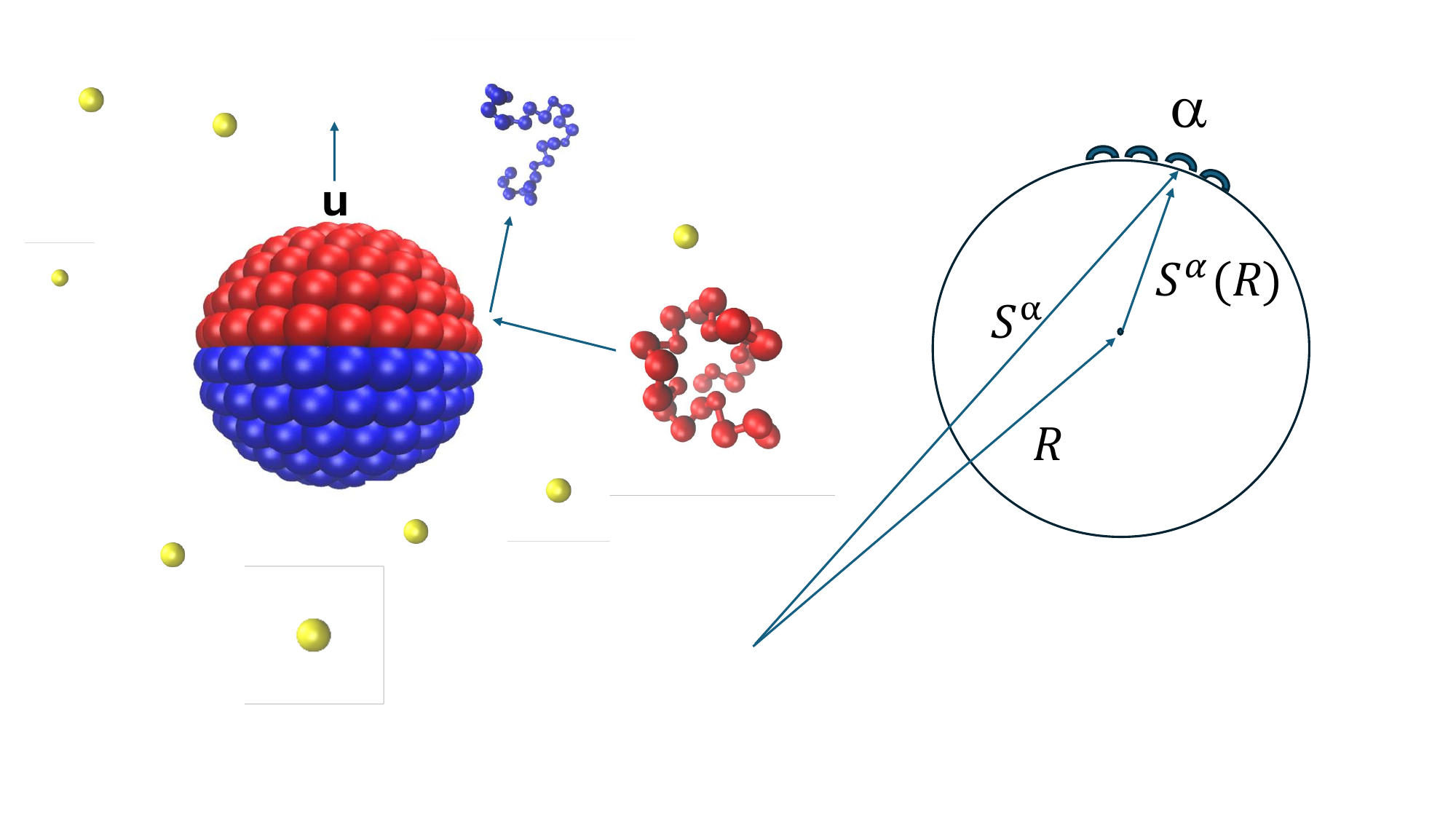}}
        \caption{ Diagram of the active Janus colloid in solution.  Two reactive molecules, in red and blue, that can interconvert when catalyzed by the catalytic beads (red) are shown. The image on the right depicts the coordinate system relative to the lab frame, where the colloid center and a site $\alpha$ on the surface of the colloid are located at positions $\bm{R}$ and $\bm{S}^\alpha$, respectively.
    } \label{fig:janus}
\end{figure}
The Janus colloid resides in a large volume of fluid far from any confining walls. It is a rigid spherical object with radius $R_J$ and a uniform mass distribution with total mass $M$ and moment of inertia $I_0$. We assume that $M \gg m$ so that the small parameter $\mu = (m/M)^{1/2}$ can be used to characterize the colloid dynamics as in derivations of Langevin dynamics for inactive~\cite{MO70} and active~\cite{RSGK20} Brownian particles. To account for its molecular structure on a coarse-grained level, the colloid is constructed from $n_s$ small spherical beads uniformly distributed on its surface.~\cite{colloid} Catalytic and noncatalytic beads are arranged in two hemispherical caps to form a Janus colloid with unit orientation vector $\bm{u}$. The surface beads, labeled by an index $\alpha$, are located at positions $\bm{S}^\alpha$ in the laboratory frame and lie at a distance $R_J$ from the center $\bm{R}$ of the colloid at locations relative to the colloid center, $\bm{S}^\alpha(\bm{R}) = \bm{S}^\alpha - \bm{R}$. The phase space coordinates of the rigid colloid are $\bm{X} = \{ \bm{R}, \bm{\theta}, \bm{P}, \bm{\Pi} \}$, where the center-of-mass momentum is $\bm{P}=M\bm{V}$, with $\bm{V}$ its velocity, and $\bm{\theta}$ is a set of Euler angles with $\bm{\Pi}$ the corresponding generalized rotational momentum. The angular momentum of the colloid, $\bm{L}$, is related to its angular velocity by $\bm{\omega}$  by  $\bm{L}=I_0 \bm{\omega}$.

Depending on their molecular configurations and bonding states, the reactive molecules can be classified as species $A$ or $B$. When interacting with the catalytic sites of the Janus colloid, the chemical reactions $A \rightleftharpoons B$ may take place due to lowering the free energy barrier that separates reactants from products. We assume that the barrier is very high in the absence of interactions with the colloid, so reactions in the fluid phase can be neglected.

The Hamiltonian for the entire system can be expressed as the sum of the kinetic energy of the colloid, $K=P^2/(2M)+L^2/(2I_0)$, and the Hamiltonian for the fluid in the presence of the fixed colloid, $H_0$, is
\begin{eqnarray*}
%\label{eq:Ham_bath_col-1}
H= K+ H_0.
\end{eqnarray*}
The Hamiltonian $H_0$ can be written as the sum of the kinetic energy of the solvent $K_S$, the Hamiltonian of the molecular species, $H_m$, the potential energy of the fluid molecules, $U_{{\rm f}}$ and the interaction energy between the colloid and fluid molecules, $U_{{\rm I}}$,
\begin{equation*}
%\label{eq:definition-ei-1}
H_0 = K_S + H_{m} + U_{{\rm f}} + U_{{\rm I}}.
\end{equation*}

The equation of motion of an arbitrary dynamical variable $B(\bm{X},\bm{x})$ of the full phase space, where $\bm{x}$ is the phase space coordinate of the solvent degrees of freedom, is
\begin{equation*}
\frac{d }{dt}B(\bm{X},\bm{x},t) = i{\cal L} B(\bm{X}, \bm{x},t),
\end{equation*}
where $i\cal{L}$ is the Liouville operator that we write as the sum of the Liouvillian operators for the colloid and fluid in the presence of the colloid, $i{\cal L} = i{\cal L}_c + i{\cal L}_0$. Further details of the interactions, Liouville operators and rigid-body dynamics are given in Appendix~\ref{app:model}.

\section{Coarse-grained variables and fluxes}\label{sec:noneq-dist}

\subsection{Coarse-grained microscopic variables}
The microscopic variables of interest are the set of colloid variables, $\bm{A}_c=(1,\bm{P},\bm{L})$, where the $1$ in this list accounts for the presence of the single colloid in the system, and the coarse-grained, slowly-varying fields for the conserved fluid variables: number $N(\bm{r})$, momentum $\bm{g}_N(\bm{r})$ and total energy ${E}(\bm{r})$ densities, as well as the reactive species densities $N_\gamma(\bm{r})$, $\gamma \in \{A,B\}$. The set of fluid variables is denoted by $\bm{A}(\bm{r})=\{N_\gamma(\bm{r}),N(\bm{r}),\bm{g}_N(\bm{r}),{E}(\bm{r}) \}$ These coarse-grained fields take the form~\cite{RKO77},
\begin{equation}
\label{eq:Delta}
A(\bm{r})=\sum_{i=1}^N A(\bm{x}_i)  \Delta(\bm{r}_{i}-\bm{r}),
\end{equation}
where $\Delta$ is a coarse-graining function that vanishes for distances $\ell_{\Delta}> \ell_{\rm int}$, where $\ell_{\rm int}$ is the range of the short-range interaction potentials. It satisfies the condition, $\int d\bm{r}' \; \Delta(\bm{r}-\bm{r}')=1$. The coarse-grained fields are related to the local microscopic fluid fields, $A_m(\bm{r})=\sum_{i=1}^N A(\bm{x}_i)  \delta(\bm{r}_{i}-\bm{r})$, by
\begin{equation}
\label{eq:A-Am-relation}
A(\bm{r})=\int d\bm{r}' \; \Delta(\bm{r}-\bm{r}')A_m(\bm{r}').
\end{equation}

The coarse-graining function $\Delta$ smooths over rapid variations appearing in microscopic density averages on short length scales $\ell_{\rm int}$.   For example, consider the gradient of the non-equilibrium average of the coarse-grained number density of particles at position $\bm{r}$ in the vicinity of a colloid at $\bm{R}$,
\begin{align}
\bm{\nabla}_{\bm{r}} n(\bm{r},t) &= \bm{\nabla}_{\bm{r}} \int d\bm{r}' \; \Delta(\bm{r} - \bm{r}') n_m(\bm{r}',t) ,
\end{align}
where $n_m(\bm{r}',t)$ is the probability density to find a solvent particle at position $\bm{r}'$ at time $t$.  For an equilibrium fluid in the presence of a colloid, the average fluid density $n_m(\bm{r}')$ is uniform except when $|\bm{r}' - \bm{R}| \lesssim \ell_{\rm int}$ and hence $\ell_{\Delta} \bm{\nabla}_{\bm{r}} n(\bm{r},t) \sim \ell_{\rm int}/\ell_{\Delta} \equiv \epsilon_\Delta$, where $\epsilon_\Delta$ is a small parameter.  The introduction of a coarse-graining function $\Delta (\bm{r})$ to define hydrodynamic densities is similar to the standard treatment of hydrodynamic equilibrium fluctuations in which one considers only smoothly varying local deviations in the microscopic densities of conserved variables.  This is accomplished by restricting the Fourier decomposition of the deviations to wave vectors $\bm{k}$ with magnitudes less than $\sim \ell_{\rm int}^{-1}$.  If no restrictions are imposed on the density fields, the microscopic densities retain rapidly varying spatial components that lead to motion on short, microscopic time scales.~\cite{coarse}  When the contributions of such Fourier components are not removed, the time scales of the evolution of the densities of conserved dynamical variables are not well-separated from those of other microscopic motions of the system, and simplifications of the dynamics to give hydrodynamic equations that are based on time-scale separation are not possible.

The expressions for the components of $\bm{A}(\bm{r})$ are as follows: The total number density of fluid molecules,
\begin{equation*}
%\label{eq:number-density}
N(\bm{r}) = \sum_{i=1}^{N}\Delta(\bm{r}_{i}-\bm{r}),
\end{equation*}
is equal to the sum of solvent and reactive solute densities, $N(\bm{r})=N_S(\bm{r})+N_R(\bm{r})$, where these densities are given by
\begin{equation*}
%\label{eq:SR-density}
N_\nu(\bm{r})= \sum_{i=1}^N \Theta^\nu_i \Delta(\bm{r}_{i}-\bm{r})
\end{equation*}
(In the text we consistently use the notation, $\gamma \in \{A,B\}$, $\nu \in \{S,R\}$ and $\lambda \in \{S,A,B\}$.) The reactive molecule density can be partitioned into the sum of the local number densities of the $A$ and $B$ species, $N_R(\bm{r})=N_A(\bm{r})+N_B(\bm{r})$. The $A$ and $B$ species variables may be defined in terms of scalar reaction coordinates $\xi_i(\bm{r}_i^{n_r})$ that are some functions of the configurational coordinates of the reactive molecules~\cite{RSGK20}, $\theta^\gamma_i(\xi_i)=\Theta_i^R H_\gamma(\xi_i(\bm{r}_i^{n_r}))$, where $H_\gamma(\xi_i(\bm{r}_i^{n_r}))$ restricts molecular configurations to species $\gamma \in \{A,B\}$: $H_A(\xi_i(\bm{r}_i^{n_r}))=H(\xi^\ddagger-\xi_i(\bm{r}_i^{n_r}))$ and $H_B(\xi_i(\bm{r}_i^{n_r}))=H(\xi_i(\bm{r}_i^{n_r})-\xi^\ddagger)$ with $H$ a Heaviside function. Thus, we can write the species densities as
\begin{equation*}
%\label{eq:A-B-densities}
N_\gamma(\bm{r})= \sum_{i=1}^N \theta^\gamma_i(\xi_i) \Delta(\bm{r}_{i}-\bm{r}).
\end{equation*}
The total momentum density of the centers of mass of the solvent and solute molecules is
\begin{equation*}
%\label{eq:momentum-density}
\bm{g}_N(\bm{r})=
\sum_{i=1}^{N}\bm{p}_{i}\Delta(\bm{r}_{i}-\bm{r})
\end{equation*}
and the total energy density $E(\bm{r})$ that includes the kinetic energy of the colloid at position $\bm{R}$ is
\begin{eqnarray*}
%\label{eq:energy-density}
E(\bm{r})&=& K \delta(\bm{r}-\bm{R}) +\sum_{i=1}^N e_i \Delta(\bm{r}_{i}-\bm{r}) \nonumber \\
&\equiv& K \delta(\bm{r}-\bm{R}) + E_N(\bm{r}) ,
\end{eqnarray*}
where $e_i$ is defined by $H_0=\sum_{i=1}^N e_i$. Note that the energy density is the only microscopic density that depends explicitly on the location and orientation of the colloid through the interaction potential $U_{\rm I}$. The entire set of variables is denoted by $\bm{C}(\bm{r})=(\bm{A}_c, \bm{A}(\bm{r}))$.

\subsection{Fluxes of colloid and fluid variables} \label{subsec:fluxes}
The fluxes of the colloid variables are $\dot{\bm{A}}_c = i{\mathcal L} \bm{A}_c= \big(0,\bm{F},\bm{T}\big)$, where $\bm{F}$ and $\bm{T}$ are the force and torque, respectively.
The fluxes of the slowly-varying fluid densities in the presence of the colloid are $\dot{\bm{A}}(\bm{r}) = i{\mathcal L} \bm{A}(\bm{r})$, and can be written as
\begin{equation}\
\label{eq:densityFlux}
\dot{\bm{A}}(\bm{r}) = \bm{J}(\bm{r}) - \nabla_{\bm{r}} \cdot \bm{j}(\bm{r}).
\end{equation}
The fluxes of the species densities are given by
\begin{equation*}
%\label{eq:Ngamma-flux}
\dot{N}_{\gamma{}}(\bm{r}) =  J_{R\gamma}(\bm{r})-\bm{\nabla{}}_r\cdot{}\bm{j}_{\gamma}(\bm{r}),
\end{equation*}
where the local reaction rate is $J_{R\gamma}(\bm{r})$ and $\bm{j}_{\gamma}(\bm{r})$ is the number density flux of species $\gamma$.
From its definition we have $\dot{\theta}^A_i(\xi_i)=-\dot{\theta}^B_i(\xi_i)$ and we can write the local reaction rate as
\begin{equation*}
%\label{eq:rate-flux-2}
J_{R\gamma}(\bm{r})\equiv \nu_\gamma J_{R}(\bm{r}),
\end{equation*}
where $\nu_\gamma$ is the stoichiometric coefficient for species $\gamma$, with $\nu_A=-1$ and $\nu_B=1$. Since we have assumed that the free energy barrier is very high in the bulk phase, this reactive flux is confined to the colloid reaction zone.

The fluxes of the other local densities are given by
\begin{align*}
%\label{eq:time-N}
\dot{N}(\bm{r}) &=-\bm{\nabla{}}_r\cdot{}\bm{g}_N(\bm{r})/m \\
%\label{eq:time-g}
\dot{\bm{g}}_N(\bm{r}) &= \bm{F}_{\rm f}(\bm{r}) -\bm{\nabla{}}_r\cdot{}\bm{\tau{}}(\bm{r}),
\end{align*}
expressed in terms of the fluid stress tensor $\bm{\tau}(\bm{r})$ and the local force on the fluid $\bm{F}_{\rm f}(\bm{r})$. We have $\int d\bm{r} \; \bm{F}_{\rm f}(\bm{r})=-\bm{F}$ since the total momentum is conserved in the absence of an external force. The local total energy flux is
\begin{align*} 
%\label{eq:Edot}
 \dot{E}(\bm{r}) &= J_E(\bm{r}) -\nabla_{\bm{r}} \cdot \bm{j}_e (\bm{r}).
\end{align*}
The total energy $\int d\bm{r} \; E(\bm{r})$ is constant so that the total integrated energy flux is zero. The full expressions for these fluxes are given in Appendix~\ref{app:flux-coarse}.

\section{non-equilibrium distributions and average fluid fields} \label{sec:local-density}

Prior to the derivation of a set of equations for the non-equilibrium averages $\bm{c}(\bm{r},t)$ of the set of slowly-varying colloid and fluid variables $\bm{C}(\bm{r})$, we describe the non-equilibrium distributions that are used to compute these average values.

The non-equilibrium distribution function for the system satisfies the Liouville equation,
\begin{equation*}
\partial_t \rho(t) =-i{\mathcal L} \rho(t),
 \end{equation*}
 so that the evolution equations for the average values follow from
 \begin{equation}
 \label{eq:fullhydro}
\partial_t \bm{c}(\bm{r},t) = \Tr[ \partial_t \rho(t) \bm{C}(\bm{r}) ] =\Tr[ \rho(t) \dot{\bm{C}}(\bm{r}) ] ,
\end{equation}
where the trace operation includes an integration over phase space and a sum over particle numbers and types,
$\Tr[\cdots]=\prod_\lambda \sum_{N_{\lambda}=0}^{\infty{}}\int d\bm{X}d\bm{x}\; \cdots$.
To derive equations for average fields $\bm{c}(\bm{r},t)$, it is useful to define a local non-equilibrium distribution $\rho_L(t)$. This distribution is constructed by maximizing the entropy functional $S(t)$,~\cite{OL79}
\begin{equation*}
S(t) = -k_B \Tr \Big[ \rho_L(t) \ln\Big( \prod_{\gamma} N_\gamma !h^{3N_\gamma} \rho_L(t) \Big) \Big],
\end{equation*}
with respect to the functional form of $\rho_L(t)$, subject to a set of constraints to be determined self-consistently:
\begin{eqnarray} \label{eq:constraints}
\bm{c}(\bm{r},t ) &=&\Tr \left[ \bm{C}(\bm{r}) \rho (t ) \right]  \nonumber \\
&=&  \Tr \left[ \bm{C}(\bm{r}) \rho_L(t) \right] \equiv \left\langle \bm{C}(\bm{r}) \right\rangle_t.
\end{eqnarray}
From Eq.~(\ref{eq:constraints}), we see that the local equilibrium distribution is constructed so that non-equilibrium averages $\bm{c}(\bm{r},t)$ of the set of slow variables $\bm{C}(\bm{r})$ at time $t$ are exactly given by their average over the normalized local equilibrium density, $\rho_L(t)$.   
%the set of local constraints defined by the normalization of the non-equilibrium distribution and non-equilibrium averages of the coarse-grained hydrodynamic densities,
We represent the set of averages as $\bm{c}(\bm{r},t) = \big( \bm{a}_c(t), \bm{a}(\bm{r},t) \big)$, where the components of $\bm{c}$ are
\begin{align*}
%\label{eq:a-components}
\bm{a}_c&(t) =\{1,\bm{P}_c (t),\bm{L}_c (t)\} \nonumber \\
\bm{a}&(\bm{r},t) =\{n_\gamma (\bm{r},t),n(\bm{r},t),m n(\bm{r},t) \bm{v}(\bm{r},t),e(\bm{r},t)\} .\nonumber
\end{align*}
Here, the average velocity of the colloid is defined by $\bm{P}_c = M \bm{V}_c$, the colloid's position can be found from $d\bm{R}_c/dt = \bm{V}_c$, and the local fluid velocity $\bm{v}(\bm{r},t)$ is defined through the non-equilibrium average of the momentum density, $\bm{g}(\bm{r},t)=m n(\bm{r},t) \bm{v}(\bm{r},t)$.

The resulting local equilibrium distribution is
\begin{equation}
\label{eq:local_eq-new}
\rho_{L}(t)=\frac{\Pi_\lambda (N_\lambda ! h^{3N_\lambda})^{-1} e^{\bm{C}(\bm{r}) \ast{} \bm{\phi}_{C}(\bm{r},t)}}{\Tr[
\Pi_\lambda (N_\lambda ! h^{3N_\lambda})^{-1}e^{\bm{C}(\bm{r})\ast \bm{\phi}_{C}(\bm{r},t)}]},
\end{equation}
expressed in terms of a set of local fields,
\begin{equation*}
\bm{\phi}_{C}(\bm{r},t)=\{\bm{\phi}_{c}(t),\bm{\phi}_{A}(\bm{r},t)\}.
\end{equation*}
In Eq.~(\ref{eq:local_eq-new}),
\begin{equation*}
\bm{C}(\bm{r})\ast{}\bm{\phi}_{C}(\bm{r},t)= \bm{A}_c \cdot \bm{\phi}_{c}+\int{}d\bm{r}\bm{A}(\bm{r})\cdot{}\bm{\phi}_{A}(\bm{r},t),
\end{equation*}
the $\ast$ denotes a scalar product or tensor contraction and integration over $\bm{r}$. The $\bm{\phi}_{C}(\bm{r},t)$ fields, which are spatial and time-dependent Lagrange multipliers that enforce the constraint conditions, are functionals of the average variables and are given by
\begin{equation*}
%\label{eq:phi_c}
\bm{\phi}_{c}(t)=\beta (-K_c,\bm{V}_c, \bm{\omega}_c),
\end{equation*}
where $K_c=P_c^2/2M+L_c^2/2I_0$, and
\begin{equation*}
%\label{eq:phi_fields}
\bm{\phi}_A(\bm{r},t) =  \big({\phi}_\gamma (\bm{r},t), {\phi}_n (\bm{r},t), {\phi}_v (\bm{r},t), \phi_e(\bm{r},t) \big),
\end{equation*}
with
\begin{eqnarray}
\label{eq:phi-fields}
&&\phi_\gamma(\bm{r},t) = \beta \tilde{\mu}_\gamma (\bm{r},t),\quad \bm{\phi}_v(\bm{r},t) = \beta\bm{v}(\bm{r},t) \\
&&\phi_n(\bm{r},t) =\beta \big( \mu_S(\bm{r},t)-\frac{1}{2} m v^2(\bm{r},t)\big), \quad \phi_e(\bm{r},t) = - \beta. \nonumber
\end{eqnarray}
The chemical potential of species $\gamma$ above is defined by $\tilde{\mu}_\gamma (\bm{r},t)={\mu}_\gamma (\bm{r},t)-{\mu}_S (\bm{r},t)$. In the more general situation where the temperature field is inhomogeneous, $\beta(\bm{r},t)$ is a function of space and time determined by the constraints.

In contrast to the case where the local equilibrium density is defined by the constrained averages of the microscopic hydrodynamic densities\cite{OL79}, the Lagrange multipliers $\bm{\phi}_A(\bm{r},t)$ are connected to the averages of the coarse grain hydrodynamic densities via non-local relations that arise from the coarse-graining function $\Delta(\bm{r})$ introduced in Eq.~(\ref{eq:Delta}). The conditions that result in Eqs.~(\ref {eq:phi-fields}) are given in Appendix~\ref{app:coarse}.

The local equilibrium distribution can also be written conveniently as a function of the instantaneous internal energy density $E^{\ddag}(\bm{r})$ in which the momenta of the fluid particles $\bm{p}_i$ are replaced by their relative momenta $\bm{p}_i^\ddag = \bm{p}_i - \langle \bm{p}_i \rangle_t = \bm{p}_i - m \bm{v}(\bm{r}_i,t)$, and the linear and angular momenta of the colloid are expressed in terms of their relative values, $\bm{P}^\ddag =\bm{P}- M\bm{V}_c$ and $\bm{L}^\ddag =\bm{L}- I_0\bm{\omega}_c$, respectively. In this case, we have
\begin{eqnarray*}
&&\bm{C}(\bm{r}) \ast \bm{\phi}_C(\bm{r},t) =
-\beta {E}^\ddag    \\
&& \quad + N_\gamma (\bm{r}) \ast \beta \tilde{\mu}_\gamma (\bm{r},t)   + N(\bm{r}) \ast  \beta \mu_S(\bm{r},t).\nonumber
\end{eqnarray*}
The non-equilibrium average of the internal energy density is $\big\langle E^\ddag (\bm{r}) \big\rangle_t = e^\ddag(\bm{r},t)$ where
\begin{equation*}
%\label{eq:internalEnergy}
e^\ddag(\bm{r},t)  = e(\bm{r},t) - \frac{1}{2} mn(\bm{r},t) v^{2}(\bm{r},t) + O(\epsilon_\Delta^2).
\end{equation*}

Since $H_0$ does not depend on the linear or angular momenta, we can define a local equilibrium distribution for the colloid kinetic terms by
\begin{equation*}
\rho^c_L(t)=e^{-\beta K^\ddag}/\int d\bm{P} d\bm{\Pi} \;e^{-\beta K^\ddag}
\end{equation*}
and write $\rho_{L}(t)=\rho^c_L(t) \rho^0_L(t)$
where
\begin{equation*}
%\label{eq:local_eq-fixed}
\rho^0_L(t)=\frac{\Pi_\lambda (N_\lambda ! h^{3N_\lambda})^{-1} e^{\bm{{A}_0}(\bm{r}) \ast{} \bm{\phi}_A(\bm{r},t)}}{\Tr'[\Pi_\lambda (N_\lambda ! h^{3N_\lambda})^{-1}e^{\bm{{A}_0}(\bm{r})\ast \bm{\phi}_{{A}}(\bm{r},t)}]},
\end{equation*}
where the set $\bm{A}_0(\bm{r})=\{N_\gamma(\bm{r}),N(\bm{r}),\bm{g}_N(\bm{r}),{E}_N(\bm{r}) \}$ differs from the set $\bm{A}$ in that the total energy density is replaced by the fluid energy density in the presence of a fixed colloid, and $\Tr'[\cdots]=\int d\bm{R} d\bm{\theta}\prod_\lambda \sum_{N_{\lambda}=0}^{\infty{}}\int d\bm{x}\; \cdots$. 

In what follows, we will assume that the conjugate fields $\bm{\phi}_{{A}} (\bm{r},t)$ are uniform sufficiently far from the location of the Janus particle $\bm{R}$, and, in particular, at the outer boundaries of the system.  Under these circumstances, averages of fluid densities $\bm{A} (\bm{r})$ at field position $\bm{r}$ with respect to the local equilibrium density are functions of the position $\bm{r} -\bm{R}$ relative to the colloid due to the translational invariance of the interaction potential between the Janus particle and the solvent. These results follow directly from the fact that
\begin{align*}
    %\label{eq:invarianceR}
    \bm{a} (\bm{r},t) &= \Tr' \sum_{i=1}^{N} \bm{a}_i (\bm{r}_i, \bm{p}_i) \Delta (\bm{r}_i - \bm{R} + \bm{R} - \bm{r}) \rho_L (t) \nonumber \\
    &= \Tr' \sum_{i=1}^N \bm{a}_i (\bm{r}_{ic}, \bm{p}_i) \Delta (\bm{r}_{ic} + \bm{R} - \bm{r}) \rho_L (t) \nonumber \\
    &= \tilde{\bm{a}}(\bm{r}-\bm{R},t).
\end{align*}
Similarly, for a fixed colloid, the relative field position $\bm{r} - \bm{R} = \bm{r}_c$ can be written in a body-fixed frame in which the Janus orientation vector $\bm{u}$ is the zenith direction and the hydrodynamic fields $\bm{a}(\bm{r},t)$ expressed in this coordinate system.

However, for a finite system, the interaction potential must include interactions of the colloid that confine it within the system's boundaries, either through an external potential that depends on $\bm{R}$ or via specific interactions with a microscopic description of the confining walls.  Such interactions break the translational symmetry of the interaction potential and result in a positional dependence of the hydrodynamic fields on the location of the colloid within the finite system.

\section{Generalized colloid and fluid equations}\label{sec:hydro-eqs}

To derive the equations of motion for the average $\bm{c}(\bm{r},t)$ fields, we first show that averages over the full density $\rho(t)$ in Eq.~(\ref{eq:fullhydro}) can be expressed as averages over the local equilibrium density $\rho_L(t)$. The time evolution of the local density satisfies
\begin{equation*} 
%\label{eq:localEvolution}
\partial_t \rho_L(t) = \rho_L(t) \tilde{\bm{C}}(\bm{r}) \ast
\dot{\bm{\phi}}_{C}(\bm{r},t ),
\end{equation*}
where $\tilde{\bm{C}}(\bm{r}) = \bm{C}(\bm{r}) - \langle \bm{C}(\bm{r}) \rangle_t$.
To relate the non-equilibrium density $\rho(t)$ to the local equilibrium density $\rho_L(t)$ that yields the same average $\bm{c}$ fields, we define the projection operators,
\begin{eqnarray*}
%\label{eq:le_proj-2}
\mathcal{P}^\dagger_{C}(t) \hat{\rho}(t) &=& \Tr[ \hat{\rho} (t) ]\rho_L(t)\\
&+& \Tr [ \hat{\rho} (t)\tilde{\bm{C}}(\bm{r}_{1})]\ast{}\langle{}\tilde{\bm{C}}\tilde{\bm{C}}
\rangle_{t}^{-1}(\bm{r}_{1},\bm{r}_{2})\ast{}\tilde{\bm{C}}(\bm{r}_{2}) \rho_L(t) \nonumber \\
\mathcal{P}_{C}(t) D &=& \langle D \rangle_t\\
&+& \langle D \tilde{\bm{C}}(\bm{r}_{1})\rangle_t \ast{}\langle{}\tilde{\bm{C}}\tilde{\bm{C}}
\rangle_{t}^{-1}(\bm{r}_{1},\bm{r}_{2})\ast{}\tilde{\bm{C}}(\bm{r}_{2}),  \nonumber
\end{eqnarray*}
where $\hat{\rho}(t)$ is a general probability density and $D$ is an arbitrary dynamical variable,
as well as their respective complements $\mathcal{Q}^\dagger_{C}(t)=1-\mathcal{P}^\dagger_{C}(t)$ and $\mathcal{Q}_{C}(t) = 1 - \mathcal{P}_{C}(t)$. These projection operators are Hermitian conjugates under the trace, $\Tr \big[ \hat{\rho}(t) \big( {\mathcal{P}}_C(t) D \big) \big] = \Tr \big[ \big( {\mathcal{P}}^\dagger (t) \hat{\rho}(t) \big) D \big]$.
Note that when $\hat{\rho}(t) = \rho (t)$, $\mathcal{P}^\dagger_{C}(t)\rho(t) = \rho_L (t)$ since $\Tr [ \tilde{C}(\bm{r}_1) \rho (t) ] = 0$.
It follows that~\cite{RSGK20}
\begin{eqnarray}\label{eq:rhob-rhoL}
&&\rho(t) = \rho_L(t) +U^\dagger_{Q_{C}}(t,0) \mathcal{Q}^\dagger_{C}(0) \rho(0)  \\
&& \qquad - \int_0^t dt_1 \; U^\dagger_{Q_{C}}(t,t_1)  \bm{{\cal F}}_{C,t_1}(\bm{r})\ast \bm{\phi}_{C}(\bm{r},t_1)  \rho_L(t_1), \nonumber
\end{eqnarray}
where $\bm{{\mathcal F}}_{C,t_1}(\bm{r})= \mathcal{Q}_{C}(t_1)i{\mathcal L} \bm{C}(\bm{r})$.  In Eq.~(\ref{eq:rhob-rhoL}), $U^\dagger_{Q_A}(t,0)$ is the projected evolution operator
\begin{equation*} 
%\label{eq:formalUdagger}
U^\dagger_{{\cal Q}_{C}}(0,t) = {\cal T}_+ \exp \left( -\int_0^t dt_1 \,
{\cal Q}^\dagger_{C}(t_1) i \Liou \right) ,
\end{equation*}
where the time-ordering operator ${\cal T}_+$ orders operators with smaller time arguments to the right of operators of larger time arguments.

Inserting Eq.~(\ref{eq:rhob-rhoL}) into Eq.~(\ref{eq:fullhydro}) and integrating by parts,
we obtain,
\begin{eqnarray}
\label{eq:av-gen-hydro}
&&\partial_t \bm{c}(\bm{r},t)  = \langle \dot{\bm{C}}(\bm{r}) \rangle_t  + \Tr[\rho(0) \bm{\mathcal{F}}_{C,t}(\bm{r},0,t)] \\
&& \qquad - \int_0^t dt_1 \; \langle \bm{\mathcal{F}}_{C,t}(\bm{r},t_1,t) \bm{\mathcal{F}}_{C,t_1}(\bm{r}') \rangle_{t_1} \ast \bm{\phi}_C(\bm{r}',t_1),\nonumber
\end{eqnarray}
where the time-dependent random force is defined as
\begin{equation*}
%\label{eq:random-force}
\bm{\mathcal{F}}_{C,t}(\bm{r},t_1,t) = U_{Q_{C}}(t,t_1)
\bm{{\mathcal F}}_{C,t_1}(\bm{r}),
\end{equation*}
and the evolution operator is given by
\begin{equation*}
U_{Q_C}(t,t_1)={\cal T}_{-} \exp \left(\int_{t_1}^t d\tau \; i\Liou {\mathcal Q}_C(\tau)  \right).
\end{equation*}
The time ordering operator ${\cal T}_{-}$ orders operators with smaller time arguments to the left of operators of larger time arguments.

Equation~(\ref{eq:av-gen-hydro}) is an exact non-local equation with memory for the non-equilibrium averages $\bm{c}(\bm{r},t)$ of the coarse-grained microscopic fields. In the subsequent development, we make approximations based on time and length scales to reduce this equation to tractable forms that apply to the physical situations of interest in this work.

While the derivation above is for average values, fluctuating fluid hydrodynamic equations may also be derived from the molecular theory, similar to those used to obtain the Langevin equations for an active colloid~\cite{RSGK20}. Such equations also have been obtained using fluctuating hydrodynamics methods~\cite{BM74,BAM77,GK18a}.

\section{Reduction to Markovian equations}\label{sec:simp}
The following analysis shows how the equations for the colloid and fluid fields may be approximated when there is a time-scale separation in the system. The variables that comprise the $\bm{C}(\bm{r})$ fields were chosen to have average values that evolve slowly in time compared to microscopic times. When the mass ratio $\mu=\sqrt{m/M} \ll 1$, the colloid linear and angular momenta vary slowly on a time scale $\tau_c$ that is much larger than microscopic times $\tau_{\rm mic}$. The coarse-graining of the microscopic fluid conserved fields smooths the interactions that occur at short intermolecular distances where strong forces act and, consequently, coarse-grained fields vary on slow hydrodynamics times $\tau_h$ which are also much larger than $\tau_{\rm mic}$. Since the $\bm{\phi}_C(\bm{r},t)$ fields are constructed to give the exact average values $\bm{c}(\bm{r},t)$, these fields vary on the above slow time scales. Thus, there are two small parameters, $\mu \sim \tau_{{\rm mic}}/\tau_c$ and $\epsilon \sim \tau_{{\rm mic}}/\tau_h$, that gauge the magnitudes of these characteristic times for the colloid and fluid fields.

This time-scale separation allows us to obtain simpler forms for the projected evolution operators that appear in the equations of motion. For any dynamical variable $B$, we can write
\begin{eqnarray*}
\mathcal{Q}_C(t_1)B &=& \mathcal{Q}_C(t)B - \int_{t}^{t_1} d\tau \, \frac{\delta \left( \mathcal{P}_C(\tau)B \right)}{\delta \bm{\phi}_C (\bm{r}' ,\tau) } * \dot{\bm{\phi}}_C(\bm{r}',\tau)\nonumber \\
&=& \mathcal{Q}_C(t)B + \mathcal{O}(\mu,\epsilon),
\end{eqnarray*}
where the second line follows from the fact that the time derivatives of the ${\bm{\phi}}_C(\bm{r},\tau)$ fields are slowly varying. In turn, using this result, we can replace the projectors $\mathcal{Q}_{C}(t_n)$ that enter the time-ordered evolution operator $U_{Q_C}(t,t_1)$
by $\mathcal{Q}_{C}(t)$, so that $U_{Q_C}(t,t_1)\approx e^{ \mathcal{Q}_{C}(t) i \mathcal{L}(t-t_1)}$.

Below, we use the following notation for the components of the average fields: for $\bm{c}(\bm{r},t)=(1,\bm{a}_c(t),\bm{a}(\bm{r},t))$ we have $\bm{a}_c(t)=(\bm{V}_c(t),\bm{\omega}_c(t))$ and $\bm{a}(\bm{r},t)=(n_\gamma(\bm{r},t),n(\bm{r},t),\bm{g}(\bm{r},t),e(\bm{r},t))$ and the indices $c= \{a_c,a\}=\{V, \omega,\gamma,n,v,e\}$.

\subsection{Active colloid equations}
The equations of motion for the colloid linear and angular momenta are simplified by using scaled variables similar to those that enter Brownian motion theory for the molecular derivation of the Langevin equation.~\cite{MO70} This scaling was also used in the molecular derivation of the Langevin equation for an active colloid.~\cite{RSGK20} The linear and angular momenta are scaled as $\mu \bm{P} = \bm{P}^*$ and $\mu \bm{L} = \bm{L}^*$, respectively, or $\mu \bm{a}_c = \bm{a}_c^*$.  The dynamics is considered in the limit where $\mu$ tends to zero while $\bm{a}_c^*$ remains finite, which implies that $\bm{a}_c \sim M^{1/2}$. Also, since $\bm{\phi}_c=(\beta \bm{V}_c,\beta \bm{\omega}_c)$, $\bm{V}_c= \bm{P}_c/M$ and $\bm{\omega}_c= \bm{L}_c/I_0$, we have $\mu \bm{\phi}_c=\mu^2 (\beta \bm{P}^*_c/m,\beta \bm{L}^*_c/I_m)\equiv\mu^2 \bm{\phi}^*_c$ where $I_m$ is the moment of inertia expressed in terms of the small mass $m$.

Taking the $a_c$ component of Eq.~(\ref{eq:av-gen-hydro}) and multiplying by $\mu$ we have
\begin{eqnarray}\label{eq:av-gen-c}
&&\frac{d}{dt} \bm{a}_c^*(t)  = \mu\langle \dot{\bm{A}}_c \rangle_t  +\mu \Tr[\rho(0) \bm{\mathcal{F}}_{c,t}(0,t)] \nonumber \\
&& \qquad \quad - \mu \int_0^t dt_1 \;  \langle \bm{\mathcal{F}}_{c,t}(t_1,t) \bm{\mathcal{F}}_{{A},t_1}(\bm{r}') \rangle_{t_1} \ast \bm{\phi}_A(\bm{r}',t_1),\nonumber \\
&& \qquad \quad - \mu^2 \int_0^t dt_1 \; \langle \bm{\mathcal{F}}_{c,t}(t_1,t) \bm{\mathcal{F}}_{c,t_1} \rangle_{t_1} \cdot \bm{\phi}^*_c(t_1),\nonumber \\
&& \qquad \quad \equiv \mu \bm{\mathsf{E}}_c (t) -\mu^2 \int_0^t dt_1 \; \bm{\mathsf{\zeta}}^b_c (t_1,t) \cdot \bm{\phi}^*_c(t_1),
\end{eqnarray}
where there are terms on the right that are at least of $\mathcal{O}(\mu)$ and $\mathcal{O}(\mu^2)$. In this equation  $\bm{\mathcal{F}}_{c,t}=\mathcal{Q}_{C}(t)\dot{\bm{A}}_c= (\mathcal{Q}_{C}(t) \bm{F},\mathcal{Q}_{C}(t) \bm{T})\equiv (\bm{F}_{t},\bm{T}_{t})$. The contributions that are  $\mathcal{O}(\mu)$ to lowest order can be written as
\begin{eqnarray*}
&&\bm{\mathsf{E}}_c (t)=\langle \dot{\bm{A}}_c \rangle_t  + \Tr[\bm{\mathcal{F}}_{c,t}(0,t) \rho(0)] \\
 && \qquad - \int_0^t dt_1 \; \langle \bm{\mathcal{F}}_{c,t} U^\dagger_{Q_{C}}(t,t_1)\bm{\mathcal{F}}_{{A},t_1}(\bm{r}') \rangle_{t_1} \ast \bm{\phi}_A(\bm{r}',t_1)\nonumber\\
 && =\Tr\Big[\bm{\dot{\bm{A}}}_c \Big(\rho_L(t)  +  U^\dagger_{Q_{C}}(t,0) \mathcal{Q}^\dagger_{C}(0) \rho(0) \nonumber \\
 &&  \qquad- \int_0^t dt_1 \;   U^\dagger_{Q_{C}}(t,t_1)\bm{\mathcal{F}}_{{A},t_1}(\bm{r}')
 \ast \bm{\phi}_A(\bm{r}',t_1)\rho_L(t_1)\Big)\Big], \nonumber
 \end{eqnarray*}
where, in the second line of this equation, we moved the propagators back onto the densities.
This term also contains contributions $O(\mu^2)$ since the fluid velocity fields have terms $O(\mu)$ arising from the slip velocity discussed below. Notice that the quantity in parentheses under the trace has the same form as that in Eq.~(\ref{eq:rhob-rhoL}) that relates $\rho(t)$ to $\rho_L(t)$, except for terms of $O(\mu^2)$ due to the coupling to colloid variables. This coupling is accounted for by the $O(\mu^2)$ term in Eq.~(\ref{eq:av-gen-c}). Furthermore, we can write $\bm{{\mathcal F}}_{{A},t_1}(\bm{r})\ast \bm{\phi}_{{A}}(\bm{r},t_1)$ as
\begin{eqnarray}
\label{eq:rhob-rhoL-term}
&&\bm{{\mathcal F}}_{{A},t_1}(\bm{r})\ast \bm{\phi}_{{A}}(\bm{r},t_1) = \beta \big[{J}_{R \gamma,t_1}(\bm{r}) \ast \tilde{\mu}_\gamma(\bm{r},t_1) \nonumber\\
&&\qquad  + \bm{j}_{\gamma,t_1}(\bm{r})\ast  \bm{\nabla}_r \tilde{\mu{}}_{\gamma{}}(\bm{r},t_1)+ \bm{\tau}_{t_1}(\bm{r}) \ast  \bm{\nabla{}}_r \bm{v}(\bm{r},t_1) \nonumber \\
&&\qquad     +  \bm{F}_{\rm f,t_1} (\bm{r}) \ast \bm{v}(\bm{r},t_1)\big],
\end{eqnarray}
taking into account the conservation of the total energy. As above, in writing this equation, we denoted the components of the projected dissipative fluxes $\bm{{\mathcal F}}_{C,t_1}(\bm{r})$ by $\bm{j}_{\gamma,t_1}(\bm{r}) ={\mathcal Q}_{C} (t_1)\bm{j}_{\gamma}(\bm{r})$, etc.

Since $i\mathcal{L}=\mu i\mathcal{L}^*_c +i\mathcal{L}_0$, we may replace $\mathcal{L}$ by $\mathcal{L}_0$ to terms $\mathcal{O}(\mu)$. Then $i\mathcal{L}_0$ acting on fluid fields yields a quantity that does not depend on the linear or angular momenta, and the projectors that enter the formulation can be replaced by $Q_{A}(t)$ and $Q^\dagger_{A}(t)$ that project only onto the fluid fields. Thus, in this limit, we have
\begin{eqnarray}
\label{eq:euler-colloid}
&&\bm{\mathsf{E}}_c (t) \approx \Tr'\Big[\bm{\dot{\bm{A}}}_c \Big( \rho^0_L(t) +U^\dagger_{Q_{A}}(t,0) \mathcal{Q}^\dagger_{{A}}(0) \rho_0(0)]\nonumber \\
 &&  - \int_0^t dt_1 \;  U^\dagger_{Q_{A}}(t,t_1)\bm{\mathcal{F}}_{{A},t_1}(\bm{r}')
 \ast \bm{\phi}_A(\bm{r}',t_1) \rho^0_L(t_1)\Big)\Big], \nonumber\\
 && \equiv \Tr'[\bm{\dot{\bm{A}}}_c\rho_0(t)\big].
\end{eqnarray}
In writing this equation, we have also taken the initial condition to be given by $\rho(0)=\rho_L^c \rho_0(0)$, and from its definition one can see that the density $\rho_0(t)$ has a structure analogous to that in $\rho(t)$ in Eq.~(\ref{eq:rhob-rhoL}).

Next, we consider the memory term $\mathcal{O}(\mu^2)$ to lowest order. After the change of variables $t'=t-t_1$, and evaluation in the limit of small $\mu$ and long times $t$ on the time scale $\tau=\mu^2 t$ we find
\begin{eqnarray*}
&&\int_0^t dt_1 \; \bm{\mathsf{\zeta}}^b_c (t_1,t) \cdot \bm{\phi}^*_c(t_1)  \\
&& \quad \approx  \Big[\int_0^\infty dt' \; \langle \big( e^{\mathcal{Q}_{{A}}(t) i \mathcal{L}_0 t'}\bm{\mathcal{F}}_{c,t}\big) \bm{\mathcal{F}}_{c,t} \rangle_{t}\Big] \cdot \bm{\phi}^*_c(t)\nonumber \\
&&\equiv \bm{\mathsf{\zeta}}^b_c  \cdot \bm{\phi}^*_c(t).
\end{eqnarray*}
The quantity $\bm{\mathsf{\zeta}}^b_c$ is a friction tensor, and the superscript ``$b$" is used to indicate that it is a bare friction since the contributions from the fluid fields are projected out of the dynamics.

Returning to unscaled coordinates, the colloid equations are
\begin{equation*}
%\label{eq:av-gen-c-2}
\frac{d}{dt} \bm{a}_c(t)  = \Tr'[\bm{\dot{\bm{A}}}_c\rho_0(t)]-\bm{\mathsf{\zeta}}^b_c  \cdot \bm{\phi}_c(t).
\end{equation*}
Writing this equation in terms of its components, we have
\begin{subequations}
\begin{align}
\label{eq:av-V-omega}
M\frac{d}{dt} \bm{V}_c &=  \Tr'[\bm{F}\rho_0(t)]-\bm{\zeta}^b_{VV}  \cdot \bm{V}_c -\bm{\zeta}^b_{V\omega}  \cdot \bm{\omega}_c \\
I_0\frac{d}{dt} \bm{\omega}_c &= \Tr'[\bm{T}\rho_0(t)]-\bm{\zeta}^b_{\omega V}  \cdot \bm{V}_c -\bm{\zeta}^b_{\omega \omega}  \cdot \bm{\omega}_c.
\end{align}
\end{subequations}
The first terms on the right contain contributions that contribute to the friction coefficients and account for the diffusiophoretic force and torque, while the remaining terms involve the bare translational and rotation friction coefficients and their cross terms. For a Janus colloid with cylindrical symmetry, there is no active torque or coupling between translation and rotation, and the equation for the colloid velocity simplifies further to give
\begin{equation}
%\label{eq:av-V}
M\frac{d}{dt} \bm{V}_c = \Tr'[\bm{F}\rho_0(t)]-\bm{\zeta}^b_{VV}  \cdot \bm{V}_c.
\end{equation}
From Eqs.~(\ref{eq:rhob-rhoL-term}) and (\ref{eq:euler-colloid}) we see that $\Tr'[\bm{F}\rho_0(t)]$ depends on the fluid fields. We construct the equation for these fields in the next section.

\subsection{Fluid equations}
The fluid equations in the presence of an active colloid follow from the $\bm{a}(\bm{r},t)$ components of Eq.~(\ref{eq:av-gen-hydro}) and are given by
\begin{eqnarray}
\label{eq:av-gen-hydro-a}
&&\partial_t \bm{a}(\bm{r},t)  = \langle \dot{\bm{A}}(\bm{r}) \rangle_t  + \Tr[\rho(0) \bm{\mathcal{F}}_{A,t}(\bm{r},0,t)] \\
&& \qquad - \int_0^t dt_1 \; \langle \bm{\mathcal{F}}_{A,t}(\bm{r},t_1,t) \bm{\mathcal{F}}_{C,t_1}(\bm{r}') \rangle_{t_1} \ast \bm{\phi}_C(\bm{r}',t_1).\nonumber
\end{eqnarray}
A Markovian approximation to the integral term in Eq.~(\ref{eq:av-gen-hydro-a}) can be made under conditions discussed above where $\mu,\epsilon \ll 1$. Letting $t_1=t-\tau$, in this approximation, to leading order in the small parameters, we can write
\begin{eqnarray*}
%\label{eq:integral-term}
&&\int_0^t dt_1 \; \langle \bm{\mathcal{F}}_{A,t}(\bm{r},t_1,t) \bm{\mathcal{F}}_{C,t_1}(\bm{r}') \rangle_{t_1} \ast \bm{\phi}_C(\bm{r}',t_1)  \\
&&\qquad \quad \approx  \Big[\int_0^\infty d\tau \; \langle \bm{\mathcal{F}}_{A,t}(\bm{r},\tau) \bm{\mathcal{F}}_{C,t}(\bm{r}') \rangle_{t}\Big] \ast \bm{\phi}_C(\bm{r}',t),\nonumber
\end{eqnarray*}
where $\bm{\mathcal{F}}_{A,t}(\bm{r},\tau)=e^{\mathcal{Q}_{A}(t) i \mathcal{L}_0\tau}\bm{\mathcal{F}}_{A,t}(\bm{r})$. The time scale condition, $\tau_{\rm mic}\ll \tau_c,\tau_h$, has been used to replace $\rho_L(t_1)$ by $\rho_L(t)$ in the kernel and make a Markovian approximation on the time integral.

Writing $\bm{{\mathcal F}}_{C,t_1}\ast \bm{\phi}_{C}$ in full we have
\begin{eqnarray}
\label{eq:rhoc-term}
&&\bm{{\mathcal F}}_{C,t_1}(\bm{r})\ast \bm{\phi}_{C}(\bm{r},t_1) = \beta\big[{J}_{R \gamma,t_1}(\bm{r}) \ast \tilde{\mu}_\gamma(\bm{r},t_1)\\
&&\qquad  + \bm{j}_{\gamma,t_1}(\bm{r})\ast  \bm{\nabla}_r \tilde{\mu{}}_{\gamma{}}(\bm{r},t_1)+ \bm{\tau}_{t_1}(\bm{r}) \ast  \bm{\nabla{}}_r \bm{v}(\bm{r},t_1) \nonumber \\
&&\qquad +\sum_\alpha \bm{F}_{\rm f,t_1}^{\alpha} (\bm{r}) \ast \bm{v}^\alpha (\bm{r},t_1)\big],\nonumber
\end{eqnarray}
which is closely related to Eq.~(\ref{eq:rhob-rhoL-term}) except that ${\bm{v}}^\alpha(\bm{r},t)$ enters in place of ${\bm{v}}(\bm{r},t)$ in the last term. This field arises from coupling to the colloid and is defined by ${\bm{v}}^\alpha(\bm{r},t) = \bm{v}(\bm{r},t) -\big( \bm{V}_c(t) + \bm{\omega}_c(t) \wedge \bm{S}^\alpha(\bm{R}) \big)$ and is the velocity field relative to the site $\alpha$ in the colloid. Using the results in Appendix~\ref{app:slip} where it is shown that 
$\sum_\alpha \bm{F}_{{\rm f},t}^\alpha(\bm{r}^\prime) \ast  \bm{v}^\alpha (\bm{r},t_1) \approx {\bm{F}}_{{\rm f},t}(\bm{r}^\prime) \ast  \bm{v}_{\rm sl}(\bm{r}^\prime,t)$,
we may express the last term in Eq.~(\ref{eq:rhoc-term}) in terms of the slip velocity, where for $|\bm{r}-\bm{R}|=R_J$,
\begin{align}
\bm{v}_{\text{sl}}(\bm{r},t) &=\bm{v}(\bm{r},t)- \bm{V}_c(t) - \bm{\omega}_c (t) \wedge (\bm{r}-\bm{R}).
\label{eq:slip}
\end{align}
Since $\bm{{\mathcal F}}_{C,t_1}\ast \bm{\phi}_{C}$ has the same form as $\bm{{\mathcal F}}_{A,t_1}\ast \bm{\phi}_{A}$ in Eq.~(\ref{eq:rhob-rhoL-term}) with $\bm{v}$ replaced by the slip velocity $\bm{v}_{\rm sl}$, we write $\bm{{\mathcal F}}_{C,t_1}\ast \bm{\phi}_{C}=\bm{{\mathcal F}}_{A,t_1}\ast \bm{\phi}_{A}$, keeping in mind that the slip velocity appears in the fluid field equations. Because the slip velocity depends on $\bm{V}_c$ and $\bm{\omega}_c$, it gives rise to a term of $O(\mu)$ in the fluid equation.

With these results, the fluid hydrodynamic equation takes the form,
\begin{equation}
\label{eq:av-gen-hydro-2}
\partial_t \bm{a}(\bm{r},t)  = \bm{\mathcal{E}}(\bm{r},t)  - \bm{\mathsf{L}}_{AA}(\bm{r},\bm{r}' , t) \ast \bm{\phi}_A(\bm{r}',t),
\end{equation}
where
\begin{equation}
\label{eq:gen-onsager}
\bm{\mathsf{L}}_{AA}(\bm{r},\bm{r}' , t)=\int_0^\infty d\tau \; \langle \bm{\mathcal{F}}_{A,t}(\bm{r},\tau) \bm{\mathcal{F}}_{A,t}(\bm{r}') \rangle_{t},
\end{equation}
and $\bm{\mathcal{E}}(\bm{r},t)= \langle \dot{\bm{A}}(\bm{r}) \rangle_t$. Since the initial condition term $\Tr[\rho(0) \bm{\mathcal{F}}_{\mathsf{A},t}(\bm{r},0,t)]$ will vary on a molecular time scale because of the projection operators it contains, it can be neglected for times long compared to $\tau_{\rm mic}$ and was not included in Eq.~(\ref{eq:av-gen-hydro-2}).

Making use of the expressions for the time derivatives of the colloid and coarse-grained fluid fields in Sec.~\ref{subsec:fluxes}, we can write the components of the random forces as
\begin{equation*}
%\label{eq:proj-fluxA}
\bm{\mathcal{F}}_{a,t}(\bm{r})=\bm{j}^{(0)}_{a,t}-\bm{\nabla{}}_r\cdot{}\bm{j}^{(1)}_{a,t}(\bm{r}),
\end{equation*}
Here and below, we use the superscripts $(\ell)$, $\ell \in \{0,1\}$, to indicate whether the flux is associated with a gradient operator. The generalized fluid hydrodynamic equations in the presence of the colloid can be written as,
\begin{equation}
\label{eq:a-hydro}
\partial_t \bm{a}(\bm{r},t) = \bm{\mathcal{E}}_a(\bm{r},t)
+ \bm{\mathcal{J}}^{(0)}_a(\bm{r},t) - \bm{\nabla{}}_r\cdot \bm{\mathcal{J}}^{(1)}_a(\bm{r},t),
\end{equation}
and, using the summation convention on repeated indices, the fluxes $\bm{\mathcal{J}}^{(\ell)}_a$ are given by
\begin{eqnarray}
\label{eq:J-fluxes}
\bm{\mathcal{J}}^{(\ell)}_a(\bm{r},t)&=& -\Big[ \bm{L}^{(\ell 0)}_{a a'}(\bm{r},\bm{r}',t)\ast \bm{\phi}_{a'}(\bm{r}',t)\nonumber \\
&&+ \bm{L}^{(\ell 1)}_{a a'}(\bm{r},\bm{r}',t)\ast \bm{\nabla}_{r'} \bm{\phi}_{a'}(\bm{r}',t) \Big] ,
\end{eqnarray}
where the $\bm{L}^{(\ell \ell')}$ coefficients are defined by
\begin{eqnarray}
\bm{L}^{(\ell \ell')}_{a a'}(\bm{r},\bm{r}' , t)&=&\int_0^\infty d\tau \; \langle \bm{j}^{(\ell)}_{a,t}(\bm{r},\tau) \bm{j}^{(\ell')}_{a',t}(\bm{r}') \rangle_{t},\label{eq:Ljj}
\end{eqnarray}
with $\bm{j}^{(\ell)}_{a,t}(\bm{r},\tau)=e^{\mathcal{Q}_{A}(t)i \mathcal{L}_0 \tau}\bm{j}^{(\ell)}_{a,t}(\bm{r})$.
The full expressions for the $\bm{\mathcal{J}}^{(0)}_a(\bm{r},t)$ and $\bm{\mathcal{J}}^{(1)}_{a}(\bm{r},t)$ fluxes are given in Appendix~\ref{app:hydro-fluxes}.

Expressing Eq.~(\ref{eq:a-hydro}) in terms of its components $a=\{\gamma, n,v,e\}$, and inserting the forms for the $\bm{\mathcal{J}}^{(\ell)}_{a}(\bm{r},t)$ fluxes given in Appendix~\ref{app:hydro-fluxes}, the full expressions for the fluid hydrodynamic equations in the presence of a moving active Janus colloid are
\begin{widetext}
\begin{subequations}
\begin{eqnarray}
\partial_t n_\gamma &=& -\bm{\nabla{}}_r\cdot{}\big( n_\gamma \bm{v} \big) - \beta{L}^{(00)}_{\gamma R} \ast  {\mathcal{A}}
- \beta\bm{L}^{(00)}_{\gamma v } \ast  \bm{v}_{\rm sl} - \beta \bm{L}^{(01)}_{\gamma \gamma'} \ast \bm{\nabla{}}_{r'}  \widetilde{\mu}_{\gamma'}  - \beta \bm{L}^{(01)}_{\gamma v} \ast  \bm{\nabla{}}_{r'} \bm{v} \label{eq:solute}   \\
&& + \bm{\nabla}_r \cdot \beta \bm{L}^{(10)}_{\gamma R}\ast {\mathcal{A}}
+ \bm{\nabla}_r \cdot \beta \bm{L}^{(10)}_{\gamma v} \ast  \bm{v}_{\rm sl} + \bm{\nabla}_r \cdot \beta \bm{L}^{(11)}_{\gamma \gamma'} \ast \bm{\nabla{}}_{r'} \widetilde{\mu}_{\gamma'}  + \bm{\nabla}_r \cdot \beta \bm{L}^{(11)}_{\gamma v} \ast \bm{\nabla{}}_{r'}\bm{v} \nonumber \\
\partial_t  \rho &=&  -\bm{\nabla{}}_r\cdot \big(\rho \bm{v}\big) \label{eq:number}\\
\partial_t \big( \rho \bm{v} \big) &=& -\bm{\nabla}_{\bm{r}}\cdot \big(\rho \bm{v}\bm{v}\big) -\bm{\nabla{}}_r\cdot{}\langle{}\bm{\tau{}}^{\ddagger{}}(\bm{r})\rangle_{t}
+\langle{}\bm{F}_{\rm f}(\bm{r})\rangle_{t}  - \beta {L}^{(00)}_{v R} \ast  {\mathcal{A}}
- \beta \bm{L}^{(00)}_{v v} \ast  \bm{v}_{\rm sl} - \beta \bm{L}^{(01)}_{v \gamma'} \ast \bm{\nabla{}}_{r'}  \widetilde{\mu}_{\gamma'}    \label{eq:momentum} \\
&&  - \beta \bm{L}^{(01)}_{v v,t} \ast  \bm{\nabla}_{r'} \bm{v} + \bm{\nabla}_r\cdot \beta \bm{L}^{(10)}_{v R} \ast {\mathcal{A}}  +
 \bm{\nabla}_r\cdot \beta \bm{L}^{(10)}_{v v} \ast  \bm{v}_{\rm sl} +\bm{\nabla}_r\cdot \beta \bm{L}^{(11)}_{v \gamma'} \ast \bm{\nabla{}}_{r'} \widetilde{\mu}_{\gamma'}  +\bm{\nabla}_r\cdot \beta \bm{L}^{(11)}_{v v} \ast  \bm{\nabla{}}_{r'}\bm{v},\nonumber
\end{eqnarray}
\end{subequations}
\end{widetext}
where $\rho(\bm{r},t) = m n(\bm{r},t)$ is the total mass density, and
\begin{equation*}
{\mathcal{A}}(\bm{r}^\prime,t) = - (\tilde{\mu}_{A}(\bm{r}^\prime,t) - \tilde{\mu}_B(\bm{r}^\prime,t) )
\end{equation*}
is the chemical affinity. Under isothermal conditions, the energy field does not couple to the species density, number density, and momentum density fields and will not be needed. Note that the equations are Galilean invariant, and can be written in a Cartesian frame co-moving with the Janus colloid in which the densities are functions of the field position relative to the moving colloid $\bm{r}_c = \bm{r} - \bm{R}$, and the fluid velocity fields are $\bm{v}_c(\bm{r}_c,t) = \bm{v} (\bm{r},t) - \bm{V}_c$.  

In the next section, we analyze these equations in the co-moving frame in further detail, both in the bulk phase and the surface layer near the colloid.

\section{Fluid phase and surface hydrodynamic equations} \label{sec:fluidAndsurface}

Equation~(\ref{eq:a-hydro}), or more explicitly Eqs.~(\ref{eq:solute})-(\ref{eq:momentum}),
describe the fluid fields in the presence of an active Janus colloid
regardless of its size. In many instances, interest centers on active colloids with micrometer size,
while the solvent molecules comprising the fluid in which they reside have nanometer or smaller
dimensions. Since the range of the solvent-colloid interactions is also on sub-nanometer or nanometer scales, the radius of the colloid $R_J$ is much larger than the characteristic size of a solvent or solute
molecule, $r_s$ and the length of the colloid-solvent interaction zone,
$\ell_{\rm int}$, $R_J \gg (r_s,\ell_{\rm int})$. Although such colloids are subject to
thermal fluctuations, hydrodynamic treatments are applicable where the fluid fields interact with
the colloid through boundary conditions. In this and the following section, we show how the
boundary conditions can be deduced from the generalized hydrodynamics equations in the presence
of the colloid. In preparation for deriving such boundary conditions, we decompose the
hydrodynamic equations into bulk and surface contributions.

\subsection{Fluid phase hydrodynamic equations:} \label{sec:fluidHydro}
We first consider the fluid field hydrodynamic equations in the absence of a Janus colloid and show that we obtain the standard results. We denote these bulk-phase fields by $\bm{a}^+(\bm{r},t)$, and drop all contributions in Eqs.~(\ref{eq:solute})-(\ref{eq:momentum}) that involve the colloid to obtain
\begin{subequations}
 \begin{align}
\partial_t n^+_\gamma &= -\bm{\nabla{}}_r\cdot{}\big( n^+_\gamma \bm{v}^+ \big) \nonumber \\
& \quad \quad +\bm{\nabla}_r \cdot \beta \bm{L}^{+}_{\gamma \gamma'} \ast \bm{\nabla{}}_{r'} \widetilde{\mu}^+_{\gamma'}  \label{eq:species+} \\
\partial_t  \rho^+ &=  -\bm{\nabla{}}_r\cdot \big(\rho^+ \bm{v}^+\big) \label{eq:number+}\\
\partial_t \big( \rho^+ \bm{v}^+ \big) &= -\bm{\nabla}_{\bm{r}}\cdot \big(\rho^+ \bm{v}^+\bm{v}^+\big) -\bm{\nabla{}}_r\cdot{}\langle{}\bm{\tau{}}^{\ddagger{}}(\bm{r})\rangle_{t}\nonumber \\
&  \qquad \qquad +\bm{\nabla}_r\cdot \beta \bm{L}^{+}_{v v} \ast  \bm{\nabla{}}_{r'}\bm{v}^+.\label{eq:momentum+}
\end{align}
\end{subequations}
 All contributions come from the $\bm{L}^{(11)}_{a a'}$ coefficients, and for these coefficients, we replace the $(11)$ superscript by $+$ to indicate their values in the absence of the colloid. In addition, since the system is isotropic, we have retained only those terms that involve coupling between forces and fluxes of the same tensorial character.

 The local equilibrium averages may be approximated by averages in the homogeneous ensemble~\cite{KO88} where $\bm{r}$-dependent dynamical variables, ${f}(\bm{r})$, in the averages are replaced by their volume averages and the $\bm{r}$ dependence arises solely from the $\bm{\phi}(\bm{r},t)$ fields, $\langle{}{f}(\bm{r})\rangle_t^{+}\approx {V}^{-1} \langle{}{f} \rangle_{H}^{+}(\bm{r},t) = V^{-1} {\rm Tr} [\rho^H_L(t) {f}]$, where ${f} =\int d\bm{r} \;{f}(\bm{r})$ and
\begin{equation}
\label{eq:local_eq-homo}
\rho_{L}(t) \approx \rho_{H}(t) =\frac{\Pi_\lambda (N_\lambda ! h^{3N_\lambda})^{-1}e^{\bm{A} \cdot \bm{\phi}^+_{A}(\bm{r},t)}}{Tr[
\Pi_\lambda (N_\lambda ! h^{3N_\lambda})^{-1}e^{\bm{A} \cdot \bm{\phi}^+_{A}(\bm{r},t)}]} ,
\end{equation}
where $\bm{A} \cdot \bm{\phi}_A^+(\bm{r},t) = -\beta E^{+} + \beta \mu^+_S(\bm{r},t) N_S + \beta \tilde{\mu}^+_{\gamma}(\bm{r},t) N_{\gamma}$.
Taking $\bm{L}^+_{aa'}(\bm{r},t) \approx \bm{L}^+_{aa',H}(\bm{r},t)$, we then have
\begin{equation}
\label{eq:L-H}
\bm{L}^+_{aa',H}(\bm{r},t)=\frac{1}{V}\int_0^\infty d\tau \; \langle \bm{j}^{(1)}_{a,t}(\tau) \bm{j}^{(1)}_{a',t} \rangle_{H}(\bm{r},t).
\end{equation}
Since the $\bm{j}^{(1)}_{a',t}$ fluxes are integrated over $\bm{r}$ and $\int d\bm{r} \; \Delta(\bm{r}-\bm{r}')=1$, the fluxes in the correlation function take their microscopic values in the homogeneous ensemble.

We then have
\begin{eqnarray}
\label{eq:+viscosity}
\beta\bm{L}^+_{vv,H}(\bm{r},t )&=&(k_BTV)^{-1}\int_0^\infty d\tau \; \langle \bm{\tau}_{t}(\tau) \bm{\tau}_{t} \rangle^+_{H}(\bm{r},t)\nonumber \\
&=& \eta(\bm{r},t) \bm{\Delta}_4+\zeta(\bm{r},t) \bm{1} \otimes \bm{1},
\end{eqnarray}
where $(\bm{\Delta}_4)_{ijkl}=\delta_{ik}\delta_{jl} +\delta_{il}\delta_{jk} -(2/3) \delta_{ij}\delta_{kl}$ is the 4th rank symmetric traceless unit tensor, with $\bm{{\sf I}}$ the 4th rank symmetric unit tensor, and $\eta(\bm{r}, t)$ and $\zeta(\bm{r},t)$, are the local shear and bulk viscosities. These transport coefficients depend locally on the hydrodynamic densities $\bm{a}^+(\bm{r}, t)$ through the functional dependence of the conjugate fields $\bm{\phi}_A^+[\bm{a}^+(\bm{r},t)]$ on the average internal energy $e^\ddagger$ and the hydrodynamic densities $n_{\gamma}(\bm{r},t)$ and $n(\bm{r},t)$.   Due to the form of the homogeneous density $\rho_H(t)$, this functional dependence is the same as that of the equilibrium transport coefficients on the equilibrium bulk density and internal energy\cite{KO88}.  In almost all circumstances, the spatial and time dependence of the transport coefficients is ignored, which is perhaps justified when nonlinear effects are small.  The inclusion of such nonlinearities introduces higher order terms of the small parameter $\tau_{\rm mic}/\tau_h$ into the hydrodynamic equations.

Similarly, we can write
\begin{eqnarray*}
%\label{eq:+diffusion}
\beta\bm{L}^+_{\gamma \gamma',H}(\bm{r}, t)&=& (k_BTV)^{-1}\int_0^\infty d\tau \; \langle \bm{j}_{\gamma,t}(\tau) \bm{j}_{\gamma',t} \rangle^+_{H}(\bm{r}, t)\nonumber \\
&=&\delta_{\gamma \gamma'}\beta D_\gamma(\bm{r}, t) n^+_\gamma(\bm{r}, t)\bm{1},
\end{eqnarray*}
that defines the diffusion coefficient of species $\gamma$. Since the solute species are assumed to be dilute, we have dropped cross-diffusion contributions. When the solution is dilute, the chemical potential of the solvent $S$ may be taken to be uniform, and the chemical potential of solute species $\gamma$ can be written as $\widetilde{\mu}^+_{\gamma}(\bm{r},t)=\widetilde{\mu}^0_{\gamma}+\beta^{-1}\ln(n^+_\gamma(\bm{r},t)/n_0)$. The fluid pressure field is given by $\langle{}\bm{\tau{}}^{\ddagger{}}(\bm{r},t)\rangle^+_{H}(t)=p^+_h(\bm{r}, t) \bm{1}$. Henceforth, we neglect the spatial dependence of the transport coefficients in the bulk phase. The fluid phase hydrodynamic equations then adopt their standard forms,
\begin{subequations}
\begin{eqnarray}
\partial_t n^+_\gamma %-\bm{\nabla{}}_r\cdot{}\big(n^+_\gamma(\bm{r},t)\bm{v}^+(\bm{r},t)\big) + {{\mathcal J}}^+_\gamma(\bm{r},t)) \\
&=& -\bm{\nabla{}}_r\cdot{} \big(n^+_\gamma\bm{v}^+\big) + D_\gamma \nabla^2_r n^+_\gamma \label{eq:solute-bulk}\\
\partial_t  \rho^+&=&  -\bm{\nabla{}}_r\cdot{}\big(\rho^+\bm{v}^+\big) \label{eq:number-bulk}\\
\partial_t \bm{g}^+ &=& -\bm{\nabla{}}_r\cdot{}\big(\bm{g}^+\bm{v}^+
+\bm{{\sf P}}^+ \big) ,\label{eq:momentum-bulk}
%\partial_t {e}^+&=& -\bm{\nabla{}}_r\cdot{}\big({e}^+\bm{v}^+
%+\bm{{\mathcal J}}_e^+ \big) .\label{eq:energy-bulk}
\end{eqnarray}
\end{subequations}
where here and below we omit the $(\bm{r},t)$ arguments for simplicity when possible. The fluid pressure tensor is
\begin{align}
\label{eq:pressure-tensor}
\bm{{\sf P}}^+ &= p^+_h \bm{1} + \bm{\Pi}^+\\
 &= p^+_h \bm{1} - 2\eta(\bm{\nabla{}}_{r}\bm{v}^+)^{\rm sym} -\big(\zeta -2\eta/3 \big) \bm{\nabla{}}_{r}\cdot\bm{v}^+\bm{1},\nonumber
\end{align}
where $\left( \bm{a} \bm{b} \right)^{\text{sym}} = (\bm{a}\bm{b} + \bm{b}\bm{a})/2$.
Below we shall assume that the fluid is incompressible, $\bm{\nabla{}}_{r}\cdot\bm{v}^+=0$, although this simplification can be relaxed.

\subsection{Hydrodynamic equations in the surface layer}\label{subsec:surf-layer}
To obtain the equations in the surface layer, we assume that the radius of the active Janus colloid is orders of magnitude larger than those of the fluid particles and the length of the interaction zone, $R_J \gg r_s,\ell_{\rm int}$.  In this circumstance, we can adopt a local Cartesian coordinate frame to construct the surface hydrodynamic equations. We take the unit vector $\hat{\bm{z}}$ to be locally normal to the colloid surface and $\bm{r}_\parallel$ to be parallel to the surface. The equations in the interfacial zone can be obtained by defining the surface fields $\bm{a}^s(\bm{r},t)$ through the equations~\cite{BAM76,RBO78}, 
\begin{align*} 
%\label{eq:surfaceExcessDefinition}
\bm{a}(\bm{r},t)&=\bm{a}^s(\bm{r},t) +\theta(z-z_0) \bm{a}^+(\bm{r},t) \nonumber 
\end{align*}
and similar expressions for other surface quantities, where $z$ is the component of $\bm{r}$ along $\hat{\bm{z}}$ and $z_0$ is the location of a dividing surface. While the location of the dividing surface may be chosen in various ways, it is convenient to suppose that it is located at the outer edge of the interaction zone but within the correlation length in the normal direction of surface transport coefficients. Using these definitions in Eq.~(\ref{eq:a-hydro}), the equation of motion for the surface fields $\bm{a}^s(\bm{r},t)$ takes the form
\begin{eqnarray}
\label{eq:a-surface-2}
&&\partial_t \bm{a}^s(\bm{r},t)= \bm{\mathcal{E}}^s_a(\bm{r},t) + \bm{\mathcal{J}}^{(0)}_{a}(\bm{r},t) - \bm{\nabla}_{r}\cdot \bm{{\mathcal J}}^{(1)s}_{a} (\bm{r},t)\nonumber \\
&& \qquad \qquad - \delta(z-z_0) \hat{\bm{z}} \cdot \bm{{\mathcal J}}^+_a(\bm{r},t),
\end{eqnarray}
where $\bm{{\mathcal J}}^+_a(\bm{r},t) = - \bm{L}_{aa'}^{+}(\bm{r},\bm{r}',t)\ast \bm{\nabla}_{{r}'} \phi_{a'}^+(\bm{r}',t)$. The surface Euler term is given by $\bm{\mathcal{E}}^s_a(\bm{r},t)=\bm{\mathcal{E}}_a(\bm{r},t) -\theta(z-z_0)   \bm{\mathcal{E}}^+_a(\bm{r},t)$, where $\bm{\mathcal{E}}^+_a$ is the Euler contribution in the bulk phase whose components are given in Eqs.~(\ref{eq:solute-bulk})-(\ref{eq:momentum-bulk}).
The surface fluxes $\bm{{\mathcal J}}^{(1)s}_{a}$ are given by
\begin{align}
\label{eq:Jtilde-full}
&\bm{\mathcal{J}}^{(1)s}_{a}(\bm{r},t) = \bm{\mathcal{J}}^{(1)}_{a}(\bm{r},t) -\theta(z-z_0) \bm{\mathcal J}^{+}_{a} (\bm{r},t).
%\bm{L}^{+}_{a a'}(\bm{r}, \bm{r}',t) \ast\bm{\nabla}_{r'} \bm{\phi}^+_{a'}(\bm{r}',t). 
\end{align}
These equations describe the evolution of the surface excess densities and their coupling to the fluxes of the bulk density fields at the interface at $z_0$.

Since the surface excess densities are discontinuous functions of $z$, we integrate them over the boundary layer corresponding to the coarse-grained fluxes, 
$\bm{a}^s_0(\bm{r}_\parallel ,t) = \int dz \; \bm{a}^s(z, \bm{r}_\parallel ,t)$, to obtain
\begin{align}
\label{eq:a-surface-int}
\partial_t \bm{a}^s_0 &(\bm{r}_\parallel,t)= \bm{\mathcal{E}}^s_{a0}(\bm{r}_\parallel,t) + \bm{\mathcal{J}}^{(0)}_{a0}(\bm{r}_\parallel,t) \nonumber \\
&  -\bm{\nabla}_{r_\parallel}\cdot \bm{{\mathcal J}}^{(1)s}_{a0} (\bm{r}_\parallel,t)
-\hat{\bm{z}} \cdot \bm{{\mathcal J}}^+_a(z_0,\bm{r}_\parallel,t),
\end{align}
where ${\bm{\mathcal J}}^{(1)s}_{a0} (\bm{r}_\parallel ,t) = \int dz \; {\bm{\mathcal J}}^{(1)s}_{a0} (z,\bm{r}_\parallel ,t)$, etc. In Appendix~\ref{app:surf-flux} we show that the fluxes in this equation can be evaluated to give
\begin{eqnarray}
\label{eq:J0-fluxes-local2-H}
\bm{\mathcal{J}}^{(0)}_a(\bm{r},t)&\approx& -\Big[ \bm{L}^{(0 0)}_{a a',H}(\bm{r}_\parallel,t)\cdot \bm{\phi}_{a'}^+(z_0,\bm{r}_\parallel,t)\nonumber \\
&+&  \bm{L}^{(0 1)}_{a a',H}(\bm{r}_\parallel,t)\cdot \bm{\nabla}_{r_\parallel} \bm{\phi}_{a'}^+(z_0,\bm{r}_\parallel,t)\Big],
\end{eqnarray}
\begin{eqnarray}
\label{eq:J1-fluxes-local-H}
&&\bm{\mathcal{J}}^{(1)s}_{a0}(\bm{r}_\parallel,t) \approx -\Big[
\bm{L}^{(1 0)}_{a a',H}(\bm{r}_\parallel,t)\cdot \bm{\phi}_{a'}^+(z_0,\bm{r}_\parallel,t)  \nonumber \\
&& \quad +  \bm{L}^{(1 1)}_{a a',H}(\bm{r}_\parallel,z_0,t)\cdot \big(\bm{\nabla}_{r_\parallel} \bm{\phi}_{a'0}^s(\bm{r}_\parallel,t)+\hat{\bm{z}} \bm{\phi}^+_{a'}(z_0,\bm{r}_\parallel,t)\big) \nonumber \\
&& \quad +  \bm{L}^{\theta}_{a a',H}(\bm{r}_\parallel,t)\cdot \bm{\nabla}_{r_\parallel} \bm{\phi}_{a'}^+(z_0,\bm{r}_\parallel,t) \Big].
\end{eqnarray}

Equation~(\ref{eq:a-surface-int}), along with the definitions of the surface fluxes in Eqs.~(\ref{eq:J0-fluxes-local2-H}) and (\ref{eq:J1-fluxes-local-H}), constitute the basic set of equations for the surface fields that describe the dynamics in the boundary layer and will be used below to derive the boundary conditions. They account for correlations in the surface layer and between the surface fields and bulk fluid fields, depending only on the properties of the coarse-grained fields and the associated small parameters that allow a Markov approximation to be made. 

In the following sections, we show how these equations can be used to obtain boundary conditions consistent in structure with continuum theories after approximations are made. Consequently, the calculations provide microscopic correlation function expressions for the parameters in these boundary conditions.

\section{Linearized surface equations}
\label{sec:linearized-surface}
The surface hydrodynamic equations derived above are local in space but are nonlinear since the homogeneous averages depend on $\bm{\phi}_a(\bm{r},t)$ fields. In the linear regime, these local equilibrium averages in the homogeneous ensemble can be replaced by equilibrium averages, and we present these linearized results below since they can be used for comparisons with standard phenomenological boundary conditions.

The general surface equations given in Eq.~(\ref{eq:a-surface-int}), with the fluxes given by Eqs.~(\ref{eq:J0-fluxes-local2-H}) and (\ref{eq:J1-fluxes-local-H}), include all couplings of the surface and bulk densities. Some of these couplings occur through transport coefficients of different tensorial character that vanish for equilibrium systems with specific symmetries.  Other couplings, such as the coupling between the reaction and the diffusive flux or the reaction and the fluid stress, are likely to be small.  Here, we consider a Janus motor system with cylindrical symmetry and neglect coupling between the reaction and diffusive transport and the reaction and fluid stress by setting $\bm{L}_{\gamma \gamma'}^{(01)} = 0$, $\bm{L}_{\gamma v}^{(01)}=0$, and $\bm{L}_{v\gamma}^{(10)} = 0$.  In the following, we also neglect the $\bm{L}^\theta_{aa'}$ transport coefficients that account for the difference between transport at the dividing surface and in the bulk.  If $z_0$ is chosen to be larger than the solvent-colloid interaction distance, these terms are small. However, they can be retained to give higher-order gradient corrections to the boundary conditions we obtain in Sec.~\ref{sec:boundary-condt}.

Written in terms of components, the linearized surface hydrodynamic equations for the species density and momentum fields are
\begin{subequations}
\begin{align}
\partial_t n_{\gamma 0}^s &=- \bm{\nabla}_{\bm{r}_\parallel}\cdot \bm{j}^s_{\gamma 0, {\rm ad}} - \bm{\nabla}_{\bm{r}_\parallel}\cdot \bm{{\mathcal J}}^{(1)s}_{\gamma 0} \label{eq:solute-surface-hyd-2} \\
& \quad + {\mathcal{J}}^{(0)}_{\gamma 0} - \hat{\bm{z}} \cdot \big( n^+_\gamma \bm{v}^+
+\bm{{\mathcal J}}^+_\gamma \big)(z_0,\bm{r}_\parallel,t) \nonumber \\
\partial_t \rho_{0}^s &= -\bm{\nabla}_{\bm{r}_\parallel} \cdot \bm{j}^s_{\rho 0,{\rm ad}}(\bm{r}_\parallel ,t) \nonumber \\
& \quad - \hat{\bm{z}} \cdot \big( \rho^+ \bm{v}^+ \big) (z_0,\bm{r}_\parallel,t)\label{eq:density-surface-hyd-2}\\
\partial_t \bm{g}_0^s &=-\bm{\nabla}_{\bm{r}_\parallel}\cdot \bm{j}^s_{v0, {\rm ad}}
-\bm{\nabla}_{\bm{r}_\parallel}\cdot \bm{{\sf P}}^s_{0}
+\langle{}\bm{F}_{\rm f0}\rangle_{t} \label{eq:mom-surface-hyd-2} \\
&\quad +\bm{\mathcal{J}}^{(0)}_{v0}(\bm{r}_\parallel ,t)
-\hat{\bm{z}} \cdot \bm{{\sf P}}^+(z_0,\bm{r}_\parallel ,t),\nonumber
\end{align} 
\end{subequations}
where the bulk phase diffusion flux is $\bm{{\mathcal J}}^+_{\gamma}= -\beta \bm{L}^{+}_{\gamma \gamma'} \cdot \bm{\nabla{}}_{r} \widetilde{\mu}^+_{\gamma'}$, the surface pressure tensor is $\bm{{\sf P}}^s_0(\bm{r}_\parallel ,t)=p_\parallel(\bm{r}_\parallel ,t) \bm{1}_2 + \bm{{\mathcal J}}^{(1)s}_{v0}$, 
and the parallel pressure is defined by $p_\parallel(\bm{r}_\parallel ,t) \bm{1}_2=   \langle \bm{\tau}_{\parallel ,\parallel,0}^{\ddagger}(\bm{r}_\parallel)\rangle_{t}^s$.

Using the results provided in Appendices~\ref{app:hydro-fluxes} and \ref{app:surf-flux}, the surface fluxes that enter these equations are as follows: The flux ${\mathcal J}^{(0)}_{\gamma 0}(\bm{r}_\parallel,t)$ can be used to define the reaction rate ${\mathcal R}$ by ${\mathcal J}^{(0)}_{\gamma 0}(\bm{r}_\parallel,t)\equiv \nu_\gamma {\mathcal R}(z_0,\bm{r}_\parallel,t)$, and is given by
\begin{eqnarray} 
\label{eq:J0gammasH-2}
{\mathcal{J}}^{(0)}_{\gamma 0}(\bm{r}_\parallel,t) &\approx&   -\beta {L}^{(00)}_{\gamma \gamma',\rm{eq}} \widetilde{\mu}^+_{\gamma'}(z_0,\bm{r}_\parallel,t).\\
&\equiv & - \nu_\gamma \beta {L}_{R}(\bm{r}_\parallel)  {\mathcal A}(z_0,\bm{r}_\parallel,t),  \nonumber
\end{eqnarray}
where 
\begin{equation}
\label{eq:react-flux}
\beta {L}_R(\bm{r}_\parallel)= H_c(\bm{r}_\parallel)\frac{\beta}{A}\int_0^\infty d \tau \; \langle J_{R,t}(\tau) J_{R,t}\rangle_{\rm{eq}},
\end{equation}
is the standard reactive flux correlation function for the reaction rate coefficient~\cite{KCM98}, and the Heaviside function $H_c(\bm{r}_\parallel)$ restricts reactions to the catalytic hemisphere. In Appendix~\ref{app:react-flux} we show that ${\mathcal R}(z_0,\bm{r}_\parallel,t)$ can be written in terms of the forward and reverse rate coefficients per unit surface area, $\kappa_{\pm}$ as
\begin{align}
\label{eq:reactionRate}
{\mathcal R}(z_0,\bm{r}_\parallel,t) &= H_c(\bm{r}_\parallel) \big[ \kappa_+ n^+_A(z_0,\bm{r}_\parallel ,t) \nonumber \\
& \qquad -\kappa_{-} n^+_B(z_0,\bm{r}_\parallel ,t) \big].    
\end{align}

The parallel component of the diffusion flux $\bm{{\mathcal J}}^{(1)s}_{\gamma 0}$ is
\begin{align}
\label{eq:Jgamma-full-reduced}
&\bm{1}_\parallel \cdot \bm{\mathcal{J}}^{(1)s}_{\gamma 0}(\bm{r}_\parallel,t)
\approx - \bm{1}_\parallel \cdot \beta \bm{L}^{(11)}_{\gamma \gamma',{\rm eq}}(z_0) \cdot \bm{\nabla}_{r_\parallel} \widetilde{\mu}_{\gamma' 0}^{s}   \\
& \quad - \bm{1}_\parallel \cdot \beta \bm{L}^{(10)}_{\gamma v,{\rm eq}}\cdot \bm{v}^+_{{\rm sl}\parallel} - \bm{1}_\parallel \cdot \beta \bm{L}^{(11)}_{\gamma v,{\rm eq}}(z_0) : \bm{\nabla}_{\bm{r}_\parallel} \hat{\bm{z}} {v}_{z0}^s \nonumber 
%& \quad - \bm{1}_\parallel \cdot \bm{L}_{\gamma \gamma',{\rm eq}}^\theta \cdot \bm{\nabla_{\bm{r}_\parallel}} \widetilde{\mu}^+_{\gamma'} \nonumber \\
%& \quad - \bm{1}_\parallel \cdot \beta \bm{L}_{\gamma v,{\rm eq}}^\theta : \left( 
%\bm{\nabla}_{\bm{r}_\parallel} \hat{\bm{z}} {v}_z^+ + \hat{\bm{z}} \partial_z \bm{v}_{\parallel}^+ \right) \nonumber ,
\end{align}
%where the symmetric projection of a second rank tensor $\bm{a}\bm{b}$ is defined as $\left( \bm{a}\bm{b} \right)^{\text{Sym}}_{\alpha \beta} = {a}_\alpha {b}_\beta \hat{\bm{\alpha}}\hat{\bm{\beta}}  + {a}_\beta \bm{b}_\alpha$.
Here, for simplicity of notation, we have not indicated the spatial dependence of the bulk fields evaluated at the dividing surface, as well as that of the surface-averaged fields. The first term in this equation describes surface diffusion. Analogous to the development presented above for dilute solute species in the bulk phase, the surface diffusion coefficient ${D}_\gamma^s (\bm{r}_\parallel ,t)$ 
%and $D_{\gamma}^\theta$ 
can be defined as
\begin{equation*}
\beta \bm{L}^{(11)}_{\gamma \gamma',{\rm eq}} (z_0)
=\beta  {D}^s_{\gamma} \, n^{s,{\rm eq}}_{\gamma 0} \delta_{\gamma \gamma'} \bm{1}_2. 
%\beta \bm{L}^{\theta}_{\gamma \gamma'} (z_0, \bm{r}_\parallel ,t) 
%&= \beta %D^\theta_{\gamma \gamma'} (z_0,\bm{r}_\parallel ,t) \, n^+_{\gamma'}(z_0, %\bm{r}_\parallel ,t) .
\end{equation*}
The second term in Eq.~(\ref{eq:Jgamma-full-reduced}) provides a microscopic expression for diffusiophoretic coupling $\beta \bm{L}^{(10)}_{\gamma v,{\rm eq}}= -\beta \bm{L}^{d}_{\gamma v}(\bm{r}_\parallel,t)$ given by
\begin{equation}
\label{eq:Ldgav}
\bm{1}_\parallel \cdot \beta \bm{L}^{d}_{\gamma v} \cdot \bm{1}_\parallel =
\frac{\beta}{A}\int_0^\infty d\tau \; \langle \bm{j}_{\parallel, \gamma,t}(\tau) \bm{F}_{\parallel,t} \rangle_{\rm{eq}} = \beta L^{d}_{\gamma v} \bm{1}_2.
\end{equation}
In writing this equation, we used the relation between the total parallel force on the colloid and the fluid, $\bm{F}_{{\rm f}\parallel,t}=-\bm{F}_{\parallel,t}$, that follows from momentum conservation.  
The third in Eq.~(\ref{eq:Jgamma-full-reduced}) accounts for the coupling between the concentration flux and the microscopic stress tensor.
This transport coefficient vanishes in the bulk of the fluid by symmetry and will contribute higher-order gradient corrections to boundary conditions.

The parallel component of flux $\bm{{\mathcal J}}^{(0)}_{v0}(\bm{r}_\parallel ,t)$ is
\begin{align}
\label{eq:J-v-tilde-full}
\bm{1}_\parallel \cdot &{\bm{{\mathcal{J}}}}^{(0)}_{v0}(\bm{r}_\parallel ,t) 
\approx -\beta \bm{1}_\parallel \cdot \bm{L}^{(00)}_{vv,\rm{eq}}
\cdot \bm{v}^+_{{\rm sl}\parallel}   \\
& \qquad - \beta \bm{1}_\parallel \cdot \bm{L}^{(01)}_{v \gamma,\rm{eq}}\cdot \bm{\nabla}_{r_\parallel} \widetilde{\mu}^+_{\gamma }  - \beta \bm{1}_\parallel \cdot \bm{L}^{(01)}_{vv ,\rm{eq}} : \bm{\nabla}_{\bm{r}_\parallel} \hat{\bm{z}} {v}_z^+. \nonumber
%\left( \bm{\nabla}_{\bm{r}_\parallel} \hat{\bm{z}} {v}_z^+ + \hat{\bm{z}} \partial_z \bm{v}_{\parallel}^+ \right) .
%\bm{1}_\parallel \hat{\bm{z}} \left( \bm{\nabla}_{\bm{r}} \bm{v}^+ \right)^{\text{sym}}_{\parallel,z}. 
\end{align}
 The first term accounts for fluid slip, and the coefficient of sliding friction, $\lambda_s =\beta L^s_{vv}$, can be defined by the correlation function
\begin{align}
\label{eq:slip-coef}
\beta \bm{1}_\parallel \cdot \bm{L}^{(00)}_{vv,\rm{eq}} \cdot \bm{1}_\parallel  &= \frac{\beta}{A} \int_0^\infty d\tau \; \left\langle \bm{F}_{{\rm f}\parallel,t}(\tau) \bm{F}_{{\rm f}\parallel,t} \right\rangle_{\rm{eq}} \\
& \equiv \beta L^s_{vv} \bm{1}_2. \nonumber
\end{align}
The relation between the time integral of the parallel force autocorrelation function and the sliding friction coefficient, also called the slip coefficient, was previously obtained by Bocquet and Barrat\cite{BB13}.  Other ways of accounting for the partial slip of the fluid velocity at boundaries are reviewed by Camargo {\it et al.}\cite{CDDEDC18} and can be consulted for additional information on various formulas for the slip coefficient.  Note that in Eq.~(\ref{eq:slip-coef}), as a result of the expansion of the correlation function in terms of the mass ratio $\mu$, the time evolution of the force $\bm{F}_{{\rm f}\parallel,t}(\tau) = \exp\{ {\cal Q}_A(t) i\Liou_0 \tau \} {\cal Q}_A(t) \bm{F}_{{\rm f}\parallel}$ is governed by the projected evolution operator ${\cal Q}_A (t)i\Liou_0$ for a {\it fixed} colloid, as in the case of the friction coefficient in a Brownian system\cite{MO70}.  Since the colloid is fixed, the integral of the force autocorrelation function over the time argument does not vanish in the thermodynamic limit.  Furthermore, since $i\Liou_0 {\cal P}_A (t) B \sim O(\tau_{\text{mic}}/\tau_h)$, the time dependence may be approximated by the unprojected evolution in the presence of a fixed particle\cite{MO70}
\begin{align*}
\lambda_s \bm{1}_2= \frac{\beta}{A} \int_0^\infty d\tau \left\langle \left( e^{i\Liou_0 \tau} \bm{F}_{{\rm f}\parallel,t} \right) \bm{F}_{{{\rm f}\parallel,t}}  \right\rangle_{\rm{eq}} \left( 1 + O(\tau_{\text{mic}}/\tau_h) \right).
\end{align*}
To evaluate the fluid slip for a finite system with $N$ fluid particles in simulations, the system size dependence must be accounted for since the time integral of the force autocorrelation function does not have a plateau at long times but rather decreases exponentially\cite{EZ93,BHP94} for $t \gg Nm/(A\lambda_s)$.  If the full Liouville operator $i\Liou$ rather than $i\Liou_0$ is retained, care must be taken to properly evaluate the transport properties using a plateau-value calculation.

The second term in Eq.~(\ref{eq:J-v-tilde-full}) accounts for diffusiophoretic effects, and the diffusiophoretic coefficient, reciprocal to that in Eq.~(\ref{eq:Ldgav}), is defined by
\begin{eqnarray}
\label{eq:diffuso0vgam}
\beta \bm{1}_\parallel \cdot \bm{L}^{(01)}_{v \gamma,\rm{eq}} \cdot \bm{1}_\parallel  &=& -\frac{\beta}{A}\int_0^\infty d\tau \; \left\langle \bm{F}_{\parallel,t}(\tau) \bm{j}_{\parallel\gamma,t} \right\rangle_{\rm{eq}}\nonumber \\
&=& - \beta {L}^d_{v \gamma} \bm{1}_2 = \beta {L}^d_{\gamma v} \bm{1}_2.
\end{eqnarray}
The third term in Eq.~(\ref{eq:J-v-tilde-full}) accounts for the coupling between the parallel component of the force and the symmetric parallel fluid stress tensor,
\begin{align}
\label{eq:Lg-definition}
    \beta \bm{1}_\parallel \cdot \bm{L}_{vv,\rm{eq}}^{(01)} : \hat{\bm{z}} \bm{1}_\parallel &= -\frac{\beta}{A}\int_0^\infty d\tau \; \left\langle \bm{F}_{\parallel,t}(\tau) \bm{\tau}_{z ,\parallel,t} \right\rangle_{\rm{eq}} \nonumber \\
    &\equiv \beta {L}_{vv}^g \bm{1}_2.
\end{align}   
The surface transport coefficient ${L}_{vv}^g$ has previously appeared in the hydrodynamic equations in the presence of solids\cite{CDDEDC18}, and has been shown to be significant\cite{CDDCE19}.

The parallel component of the surface momentum flux $\bm{1}_\parallel \cdot \bm{\mathcal{J}}^{(1)s}_{v0}$ is
\begin{eqnarray*}
\bm{1}_\parallel \cdot \bm{\mathcal{J}}^{(1)s}_{v0} \cdot \bm{1}_\parallel 
&\approx&- \bm{1}_\parallel  \bm{1}_\parallel :2 \beta \bm{L}_{vv,{\rm eq}}^{(11)} : \big( \bm{\nabla}_{r_\parallel} \bm{v}_{0\parallel}^s \big)^{\text{sym}} .
%\nonumber \\
%&&-  \bm{1}_\parallel \bm{1}_\parallel : 2\beta\bm{L}^{\theta}_{vv,{\rm eq}} : \left( \bm{\nabla}_{\bm{r}_\parallel} \bm{v}_\parallel^+ \right)^{\text{sym}}.
\end{eqnarray*}
The correlation function expression for $\bm{L}_{vv,t}^{(11)s}$ is the surface analog of that in Eq.~(\ref{eq:+viscosity}) for the fluid shear and bulk viscosities,
\begin{eqnarray}\label{eq:surf-viscosity}
\beta \bm{L}_{vv,{\rm eq}}^{(11)}&=&
\frac{\beta}{A}\int_0^\infty d\tau \; \langle \bm{\tau}_{\parallel,\parallel,t}(\tau) \bm{\tau}_{\parallel,\parallel,t}\rangle_{\rm eq}(\bm{r}_\parallel ,t),\nonumber \\
&=& \eta^s \bm{\Delta}_2+\zeta^s \bm{1}_2 \otimes \bm{1}_2,
\end{eqnarray}
where the $\bm{\Delta}_2$ and $\bm{1}_2$ are the surface analogs of the tensors in Eq.~(\ref{eq:+viscosity}). In the second line, we defined the surface shear $\eta^s$ and bulk $\zeta^s$ viscosities.  Thus, 
\begin{align*}
\bm{\nabla}_{\bm{r}_\parallel} \cdot &\bm{\mathcal{J}}^{(1)s}_{v0} \cdot \bm{1}_\parallel \approx -\eta^s {\nabla}^2_\parallel \bm{v}_{0\parallel}^s 
% \\ & \quad 
- ( \zeta^s  - \eta^s) \bm{\nabla}_{\bm{r}_\parallel}\big(\bm{\nabla}_{\bm{r}_\parallel} \cdot \bm{v}_{0 \parallel}^s \big) .
\end{align*}
%If one assumes that the fluid flow in the surface layer is incompressible, $\bm{\nabla}_{\bm{r}_\parallel} \cdot \bm{v}^s_{0\parallel} = 0$, then $
%The $\bm{L}^{\theta}_{vv,{\rm eq}}$ coefficient has an interpretation like that in the diffusion flux and is likely to be small. 

The normal component of flux $\bm{{\mathcal J}}^{(0)}_{v0}(\bm{r}_\parallel ,t)$ is
\begin{align*}
%\label{eq:Jv0-normal}
    \hat{\bm{z}} \cdot \bm{{\mathcal J}}^{(0)}_{v0}(\bm{r}_\parallel ,t) &= - \beta L_{vv}^{z,xx} \bm{\nabla}_{\bm{r}_\parallel} \cdot \bm{v}_\parallel^+ 
%  &  -\hat{\bm{z}} \cdot \beta 2\bm{L}_{vv,{\rm eq}}^{(01)} : \big( \bm{\nabla}_{{r}_\parallel} \bm{v}^+_{\parallel} \big)^{\text{sym}},
\end{align*}
where the transport coefficient $L_{vv}^{z,xx} \bm{1}_2= \hat{\bm{z}} \cdot \bm{L}_{vv}^{(01)} : \bm{1}_\parallel \bm{1}_\parallel$ accounts for the small cross-coupling between the normal component of the force and the parallel stress.

%\begin{align*}
%\beta \hat{\bm{z}} \cdot \bm{L}_{vv}^{(01)} : \bm{1}_\parallel \bm{1}_\parallel &=
%\frac{\beta}{A} \int_0^\infty d\tau \; \left\langle \bm{j}_{z\gamma,t}(\tau) %\bm{\tau}^{\parallel \parallel}_{t} \right\rangle_{\rm{eq}} ,
%\end{align*}

The normal component of flux $\bm{1}_\parallel\cdot \bm{{\mathcal J}}^{(1)s}_{v0}(\bm{r}_\parallel ,t)$ is
\begin{align*}
%\label{eq:jv1-normal}
\bm{1}_\parallel &\cdot \bm{{\mathcal J}}^{(1)s}_{v0}(\bm{r}_\parallel ,t) \cdot \hat{\bm{z}} =
-\beta \hat{\bm{z}} \bm{1}_\parallel:\bm{L}^{(10)}_{vv,\rm{eq}}  \cdot \bm{v}_{{\rm sl}\parallel}^+ \nonumber \\
&\quad -\hat{\bm{z}} \bm{1}_\parallel :\Big[\beta \bm{L}_{v\gamma,\rm{eq}}^{(11)}(z_0) \cdot\bm{\nabla}_{{r}_\parallel}  \widetilde{\mu}^s_{\gamma 0}   
%+ \beta\bm{L}_{v\gamma,\rm{eq}}^\theta \cdot \bm{\nabla}_{{r}_\parallel}   \widetilde{\mu}^+_{\gamma} 
\nonumber \\
& \quad + \beta \bm{L}_{vv,\rm{eq}}^{(11)}(z_0) : \bm{\nabla}_{\bm{r}_\parallel} \hat{\bm{z}}{v}_{0 z}^s + \beta \bm{L}_{vv,\rm{eq}}^{(11)}(z_0) : \hat{\bm{z}} \bm{v}_\parallel^+ \Big]. 
%\nonumber \\
%& \quad + \beta \bm{L}_{v v,\rm{eq}}^\theta   :  \left( \bm{\nabla}_{{r}_\parallel} \hat{\bm{z}} {v}_z^+  + \hat{\bm{z}} \partial_z \bm{v}^+_\parallel \right)  \Big] .   
\end{align*}
We note that the term $\bm{L}^{(10)}_{vv,{\rm eq}}$ is related to the transport coefficient ${L}_{vv}^g$ defined above by
\begin{align}
\label{eq:Lh-definition}
    \beta \hat{\bm{z}} \bm{1}_\parallel : \bm{L}_{vv,\rm{eq}}^{(10)} \cdot \bm{1}_\parallel  &= -\frac{\beta}{A}\int_0^\infty d\tau \; \left\langle  \bm{\tau}_{\parallel,z,t}(\tau) \bm{F}_{\parallel,t} \right\rangle_{\rm{eq}} \nonumber \\
    &\equiv \beta {L}_{vv}^h \bm{1}_2=\beta {L}_{vv}^g \bm{1}_2,
\end{align} 
and a related quantity appears in the hydrodynamic equations in the presence of solids~\cite{CDDEDC18}. From their structures, all the transport coefficients in the expression in square brackets above are expected to be small.

\section{Boundary conditions} \label{sec:boundary-condt}

In macroscopic descriptions of an active Janus motor immersed in a non-equilibrium fluid, the Navier-Stokes or Stokes equations for hydrodynamic fluid densities are usually solved by treating the effect of a translating and rotating active colloid as a boundary condition.~\cite{A86,A89,ALP82}  The presence of the slow-moving colloid induces motion in the fluid that is established on a rapid time scale relative to changes in the colloid degrees of freedom.  In the simplest case, the fluid equations are solved subject to the instantaneous values of a uniformly translating and rotating colloid.  Based on the surface and bulk hydrodynamic equations above, in this section, we derive the boundary conditions on the concentration and velocity fields from the molecular theory. When writing the boundary conditions below, we retain only the dominant terms in the dissipative fluxes derived in the previous section.

The boundary conditions on the concentration fields may be obtained by writing Eq.~(\ref{eq:solute-surface-hyd-2}) as
\begin{equation}
\label{eq:n-gamma-simple-BC}
 D_\gamma \hat{\bm{z}} \cdot \bm{\nabla}_{\bm{r}}n^+_\gamma(\bm{r},t)\mid_{z_0}=
\nu_\gamma {\mathcal R}(z_0,\bm{r}_\parallel ,t)-\Sigma^s_{\gamma},
\end{equation}
where the diffusion surface sink term $\Sigma^s_\gamma$ is defined by
\begin{equation}
\label{eq:gamma-sink}
\Sigma^s_\gamma = \partial_t n_{\gamma 0}^s + \bm{\nabla}_{\bm{r}_\parallel}\cdot \bm{j}^s_{\gamma 0, {\rm ad}} + \bm{\nabla}_{\bm{r}_\parallel}\cdot \bm{{\mathcal J}}^{(1)s}_{\gamma 0},
\end{equation}
where
\begin{align}
\label{eq:species-parallel-BC-3}
\bm{\mathcal{J}}^{(1)s}_{\gamma 0}
\approx - \beta \bm{L}^{(11)}_{\gamma \gamma',{\rm eq}} (z_0) \bm{\nabla}_{\bm{r}_\parallel} \widetilde{\mu}^s_{\gamma' 0} + \beta {L}^{d}_{\gamma v} \bm{v}^+_{{\rm sl}\parallel} .
\end{align}

The boundary condition that follows from Eq.~(\ref{eq:density-surface-hyd-2}) is,
\begin{align*}
\hat{\bm{z}} \cdot \bm{v}^{+} (z_0, \bm{r}_\parallel ,t) \rho^+ (z_0, \bm{r}_\parallel ,t) &= -\Sigma^s_n,
\end{align*}
where the mass sink is
\begin{equation} 
\label{eq:massSink}
\Sigma^s_n = \partial_t \rho_0^s + \bm{\nabla}_{\bm{r}_\parallel}\cdot \bm{j}^s_{\rho 0,{\rm ad}}.
\end{equation}
If the mass sink is zero, we have  $\hat{\bm{z}} \cdot \bm{v}^{+} (z_0, \bm{r}_\parallel ,t)=0$.

The boundary condition on the fluid velocity can be determined from Eq.~(\ref{eq:mom-surface-hyd-2}), written as
\begin{equation*}
\hat{\bm{z}} \cdot \bm{{\sf P}}^+(z_0,\bm{r}_\parallel ,t)=  \bm{\mathcal{J}}^{(0)}_{v0}(\bm{r}_\parallel ,t) - \bm{\Sigma}_v^s,
\end{equation*}
where the momentum sink is defined to be
\begin{align*}
\bm{\Sigma}_v^s &= \partial_t \bm{g}_0^s +\bm{\nabla}_{\bm{r}_\parallel}\cdot \bm{j}^s_{v0, {\rm ad}} + \bm{\nabla}_{\bm{r}_\parallel}\cdot \bm{{\sf P}}^s_{0} -\langle{}\bm{F}_{\rm f0}\rangle_{t}.
\end{align*}

For the parallel component of the fluid velocity, using Eq.~(\ref{eq:J-v-tilde-full}) for $\bm{1}_\parallel \cdot {\bm{{\mathcal{J}}}}^{(0)}_{v0}$, we write
\begin{align}
\label{eq:momentum-parallel-BC}
\hat{\bm{z}} \cdot \bm{{\sf P}}^+(z_0,\bm{r}_\parallel ,t)\cdot \bm{1}_\parallel &= -\beta L^s_{vv} \bm{v}^+_{{\rm sl}\parallel} - \beta {L}_{vv}^{g} \bm{\nabla}_{\bm{r}_\parallel} v_{z}^+ \nonumber \\
%\left( \bm{\nabla}_{\bm{r}} \bm{v}^+ \right)^{\text{sym}}_{\parallel,z} \nonumber \\
& + \beta L^{d}_{v \gamma}\bm{\nabla}_{\bm{r}_\parallel} \widetilde{\mu}^+_{\gamma } - \bm{\Sigma}_{v \parallel}^s,
\end{align}
where the parallel momentum sink term reads
\begin{eqnarray} 
\label{eq:momentumSink}
&&\bm{\Sigma}_{v\parallel}^s = \bm{\nabla}_{\bm{r}_\parallel} p_\parallel 
-\langle{}\bm{F}_{{\rm f}\parallel 0}\rangle_{t} +\partial_t \bm{g}_{\parallel 0}^s +\bm{\nabla}_{\bm{r}_\parallel}\cdot \bm{j}^s_{v0, {\rm ad}} \nonumber  \\
&& \qquad  -\eta^s {\nabla}^2_{\bm{r}_\parallel} \bm{v}_{0\parallel }^s - ( \eta^s -  \zeta^s ) \bm{\nabla}_{\bm{r}_\parallel} \big( \bm{\nabla}_{\bm{r}_\parallel} \cdot \bm{v}_{0 \parallel}^s \big).
\end{eqnarray}

Equations~(\ref{eq:n-gamma-simple-BC}) and (\ref{eq:momentum-parallel-BC}) are the general boundary conditions for the species concentration and fluid velocity fields. If surface flow terms vanish~\cite{RKO77,RBO78,RO83}, the boundary conditions simplify to
\begin{subequations}
\begin{align}
\hat{\bm{z}} \cdot \bm{v}^+ &= 0 \\
\hat{\bm{z}} \cdot \bm{{\sf P}}^+(z_0)\cdot \bm{1}_\parallel &=
-\beta L^s_{vv} \bm{v}^+_{{\rm sl}\parallel}  + \beta L^{d}_{v \gamma}  \bm{\nabla}_{ \bm{r}_\parallel} \widetilde{\mu}^+_{\gamma}  \nonumber \\
&\qquad  - \beta L_{vv}^g   \bm{\nabla}_{\bm{r}_\parallel} \bm{v}_z^+  \label{eq:momentum-parallel-BC-3} \\
 D_\gamma \hat{\bm{z}} \cdot \bm{\nabla}_{\bm{r}}n^+_\gamma(\bm{r},t)\mid_{z_0} &=
\nu_\gamma {\mathcal R}(z_0,\bm{r}_\parallel ,t) \nonumber \\
&\qquad - \beta L_{\gamma v}^d \, \bm{\nabla}_{\bm{r}_\parallel} \cdot \bm{v}_\parallel^+ .
\label{eq:species-BC} 
\end{align}
\end{subequations}
Generally, only the leading order terms in the gradients of the couplings to each field are retained when applying the boundary conditions. Equation~(\ref{eq:species-BC}) is a generalization of the radiation boundary condition applied to reactions, where the intrinsic reaction rate coefficients $\kappa_{\pm}$ that appear in the reaction rate $\mathcal{R}(z_0,\bm{r}_\parallel, t)$ given in Eq.~(\ref{eq:reactionRate}) depend on the reactive processes within the boundary layer, while the diffusive flux term accounts for the concentration gradients in the vicinity of the colloid, which have a much longer range. Both of these contributions are controlled by the effective reaction radius. Given that the boundary layer is small on the scale of the colloid size $R_J$, the dependence on $z_0$ is weak, provided it is in the vicinity of the boundary layer.

If we neglect the small dissipative contributions to the normal component of the fluid velocity equation, we find
\begin{align}
\label{eq:normal-pressure}
\hat{\bm{z}} \cdot \bm{{\sf P}}^+(z_0,\bm{r}_\parallel ,t) \cdot \hat{\bm{z}} = p_h^+(z_0) &=\hat{\bm{z}} \cdot\langle{}\bm{F}_{{\rm f}0}\rangle_{t}(\bm{r}_\parallel ,t) \nonumber \\
&\quad + \beta L_{vv}^g \bm{\nabla}_{\bm{r}_\parallel} \cdot \bm{v}_{\parallel}^+,
\end{align}
which relates the fluid pressure at $z_0$ for an incompressible fluid to the normal component of the average force integrated over the surface layer.

%These boundary conditions are consistent with those derived from non-equilibrium thermodynamics~\cite{GK18a,GK18b,GK19}. Specifically, the pair of equations~(\ref{eq:momentum-parallel-BC-3}) and (\ref{eq:species-parallel-BC-3}) have the same structure as those in Eqs.~(15) and (16) in Ref.~[\onlinecite{GK18a}] where surface viscous flows were also neglected. However, now we have microscopic correlation function expressions for the transport coefficients given in Eqs.~(\ref{eq:slip-coef})-(\ref{eq:surf-viscosity}). 
%In addition, the correlation functions involve averages in the non-equilibrium homogeneous ensemble, so they are valid beyond linear irreversible thermodynamics.

\subsection*{Solution of the Janus system with uniform interactions}

The boundary conditions in Eqs.~(\ref{eq:momentum-parallel-BC-3}) and (\ref{eq:species-BC}) have forms that are similar to those in Eqs.~(15) and (16) of Ref.~[\onlinecite{GK18a}], and are structurally identical if the velocity-dependent term in Eq.~(\ref{eq:species-BC}), which is higher order in the P\'eclet number, is neglected.  The additional coupling term involving $L_{vv}^g$ in Eq.~(\ref{eq:momentum-parallel-BC-3}) adds a correction to the fluid viscosity without changing the form of the boundary conditions. In this circumstance, we may directly use the induced force methods\cite{BM74} to solve the fluid Navier-Stokes equations subject to the boundary conditions to determine friction and diffusiophoretic force that appear in the equation of motion for the colloid average velocity:
\begin{equation*}
M\frac{d}{dt} \bm{V}_c= \bm{F}_d -\zeta_t \bm{V}_c,
\end{equation*}
where the translational friction coefficient is
\begin{equation*}
\zeta_t= 6 \pi \eta R_J \frac{(1+2b/R_J)}{(1+3b/R_J)},
\end{equation*}
and the diffusiophoretic force is
\begin{equation*}
\bm{F}_d= \frac{6 \pi \eta R_J}{(1+3b/R_J)}  \frac{\beta L^d_{v \gamma}}{\lambda_s} \overline{\bm{\nabla}_{\bm{r}_\parallel} \mu_\gamma}^s,
\end{equation*}
where $b=(\eta - \beta L^g_{vv})/\lambda_s$ and the overline denotes an average over the sphere's surface. The correlation function expressions for the slip coefficient $\beta L^s_{vv}=\lambda_s$, the diffusiophoretic coupling $\beta L^d_{v \gamma}$ and $\beta L_{vv}^g$, are given in Eqs.~(\ref{eq:slip-coef}), (\ref{eq:diffuso0vgam}), and (\ref{eq:Lg-definition}), respectively.

\section{Discussion and Conclusions} \label{sec:conclusions}
The general expressions for the transport coefficients $\bm{L}_{a a'}^{(\ell \ell')}(\bm{r},\bm{r}')$ in Eq.~(\ref{eq:Ljj}) are nonlocal functions of the spatial coordinates $\bm{r}$ and $\bm{r}'$  dependent on the choice of the coarse-graining function $\Delta (\bm{r})$ that smooths over short spatial variations of length scale $\ell_{\text{int}}$.  The coarse-graining procedure is necessary to define the set of hydrodynamic fields whose evolution is slow compared to microscopic time scales.  In turn, the set of slow density fields $\bm{A} (\bm{r})$ determines the form of the dissipative flux $\bm{j}^{(\ell)}_a(\bm{r})$ that appears in the transport coefficients, and, consequently, the form of the coupling of the hydrodynamic densities.  For example, as has been noted previously\cite{CDDEDC18}, if the set of slow variables includes the microscopic number density $N_m(\bm{r})$, then one finds that $\bm{j}_v^{(0)}(\bm{r}) = 0$ and the slip coefficient $\lambda_s = 0$. However, including short-length scale variations in the density results in the coupling of the velocity field to rapid motions of the average density field $n(\bm{r},t)$, which evolves on a molecular time scale.  Under these circumstances, the equations of motion of the velocity and density fields are nonlocal, both spatially and temporally.  The coarse-graining procedure, or a restriction in the Fourier components of the transformed densities, is a minimal requirement to obtain Markovian equations of motion. This requirement precludes the possibility of constructing a Markovian dynamic density functional theory.  Nonetheless, the spatial dependence of both the bulk and surface transport coefficients simplifies in the Navier-Stokes limit in which all terms of order $\epsilon_\Delta$ are dropped in their evaluation.

The precise nature of the symmetries of the transport coefficients in the general hydrodynamic equations for the fluid fields, Eqs.~(\ref{eq:solute})-(\ref{eq:momentum}), depends on the microscopic details of the solvent-colloid interaction potential.  For example, if each type of solvent particle interacts with all sites on the colloid with the same potential, then any inhomogeneity in the concentration field of the reactive species produced by the asymmetric catalytic activity on the surface of the Janus particle does not change the interaction energy.  Consequently, as was assumed in Sec.~\ref{sec:linearized-surface}, the transport coefficients are axisymmetric around the normal axis $\hat{\bm{z}}$ and second-rank tensors, such as the surface friction tensor $\bm{L}^{(00)}_{vv}$, have two non-zero components, $\bm{L}_{vv}^{z,z} = \hat{\bm{z}} \cdot \bm{L}^{(00)}_{vv} \cdot \hat{\bm{z}}$ and $\lambda_s={L}_{vv}^{\parallel , \parallel} = \bm{1}_{\parallel} \cdot \bm{L}^{(00)}_{vv} \cdot \bm{1}_\parallel$, and a third rank tensor $\bm{L}_{vv}^{(10)}$ with a symmetric microscopic stress tensor has three non-zero components, $\bm{L}^{\parallel \parallel , z}_{vv}$, $\bm{L}^{\parallel z,\parallel}_{vv}$, and $\bm{L}_{vv}^{zz,z}$.  In the general case in which the reactive species interact differently with sites on opposing hemispheres, the surface friction tensor and the coupling tensor $\bm{L}_{vv}^{(10)}$ have $9$ and $27$ components, respectively\cite{CDDCE19}.  However, in the dilute solution limit, $n_\gamma (\bm{r},t) \ll n(\bm{r},t)$, the non-isotropic components of the tensors are expected to be small provided the solvent-site interaction potential is the same for all sites.

Generally, the local equilibrium density $\rho^0_L (t)$ is neither even nor odd under time-reversal due to its dependence on the velocity field, $\bm{v}(\bm{r},t)$.  Consequently, the non-equilibrium transport coefficients do not obey the Onsager reciprocal relations that hold in equilibrium\cite{O31a,O31b},
\begin{align} \label{eq:onsager}
\bm{L}_{a a'}^{(\ell \ell')}(\bm{r},\bm{r}') &= \eta_a \eta_{a'} \bm{L}_{a' a}^{(\ell' \ell)} (\bm{r}',\bm{r}),
\end{align}
where $\eta_a = \pm 1$ is the signature of $a$ under time-reversal.  However, the relations do hold to Navier-Stokes order in the hydrodynamic equations.  Neglecting gradients of the hydrodynamic fields, the local equilibrium density $\rho^0_L (\bm{r},t)$ can be approximated in ensemble averages of densities in the region $\bm{r}$ by the homogeneous density $\rho_H(\bm{r},t) = \exp \left\{\bm{A} \cdot \bm{\phi}_A(\bm{r},t) \right\} /Z_H(\bm{r},t)$, where $Z_H(\bm{r},t)$ ensures normalization, in which the conjugate fields $\phi_A(\bm{r},t)$ are fixed and uniform. For this density, we may write $\bm{A} \cdot \bm{\phi}_A (\bm{r},t) = -\beta E^\ddagger(t) +\beta \mu_S(\bm{r},t) N + \beta \tilde{\mu}_\gamma (\bm{r},t) N_\gamma $, where the energy is expressed in terms of the relative momentum $\bm{p}_i^\ddagger = \bm{p}_i - m \bm{v}(\bm{r}_i, t)$.  The Liouville operator for the fluid fields in the presence of the fixed colloid can be written as $i\Liou_0 = i\Liou_0^\ddagger + \sum_i \bm{v}(\bm{r}_i,t) \cdot \bm{\nabla}_{\bm{r}_i} =i \Liou_0^\ddagger + i\Liou_v$, where $i\Liou_0^\ddagger$ is the Liouville operator with the momenta $\bm{p}_i$ replaced by the relative momenta $\bm{p}_i^\ddagger$, and the evolution operator $U_A(\tau) = \exp\{ {\cal{Q}}_A(t) i\Liou_0 \tau \}$ expanded to get
\begin{align}
U_A(\tau)  &= U_A^{+}(\tau) 
 \\ 
& + \int_0^\tau d\tau' \, U_A^+(\tau - \tau')  {\cal Q}_A(t) i\Liou_v U_A^+ (\tau') + \dots ,\nonumber
\end{align}
where $U_A^+(\tau) = \exp\{ {\cal{Q}}_A(t) i\Liou_0^\ddagger \tau \}$.  Since the projected dynamics has no slow time dependence (ignoring the pronounced mode-coupling effects observed in non-equilibrium systems\cite{MO82,SO94}), time correlation functions of the form $\left\langle [U_A^+(\tau) \bm{j}(\bm{r}) ] \bm{j}(\bm{r}') \right\rangle_t$ vanish for $\tau \gg \tau_{\text{mic}}$, which implies that
\begin{align*}
&\left\langle [U_A (\tau) \bm{j}(\bm{r}) ] \bm{j}(\bm{r}') \right\rangle_t =
\left\langle [U_A^+(\tau) \bm{j}(\bm{r}) ] \bm{j}(\bm{r}') \right\rangle_H  \\
& - \bm{v}(\bm{r},t)\tau_{\text{mic}}  \cdot \nabla_{\bm{r}} \left\langle [U_A^+(\tau) \bm{j}(\bm{r}) ] \bm{j}(\bm{r}') \right\rangle_H  + \dots  \\
&= \left\langle [U_A^+(\tau) \bm{j}(\bm{r}) ] \bm{j}(\bm{r}') \right\rangle_H + O(\bm{v}(\bm{r},t)\tau_{\text{mic}}/\ell_{\Delta}),
\end{align*}
where terms of order $\tau_{\text{mic}} \nabla_{\bm{r}} \bm{v}(\bm{r},t)$ have been dropped.  The correlation function $\left\langle [U_A^+(\tau) \bm{j}(\bm{r}) ] \bm{j}(\bm{r}') \right\rangle_H$ is independent of the velocity field $v(\bm{r},t)$ in the homogeneous ensemble\cite{KO88} since the integration over the momenta $\bm{p}_i$ in the ensemble average can be replaced by an integration over the relative momenta $\bm{p}_i^\ddagger$.  In this limit, the transport coefficients at position $\bm{r}$ can be evaluated in an equilibrium-like ensemble with the chemical potentials fixed at their values at position $\bm{r}$ and time $t$. Since the homogeneous density $\rho_H(\bm{r},t)$ obeys $i\Liou_0^\ddagger \rho_H (\bm{r},t) = 0$ and is invariant under the time inversion operator ${\cal T} G(\bm{r}_i, \bm{p}^\ddagger_i, t) = G(\bm{r}_i, - \bm{p}_i^\ddagger, -t)$, the transport coefficients obey Eq.~(\ref{eq:onsager}).

Our development of the surface hydrodynamic equations holds for any choice of the spherical dividing surface at $z_0$ between the colloid and the fluid.  As is the case in the equilibrium thermodynamics of surfaces\cite{Gibbs1878,RW02,SGKOD12,ROP22},   the value of $z_0$, and thus the excess surface densities $\bm{a}^s(\bm{r},t)$, can be chosen in a number of different ways. Nonetheless, to derive boundary conditions for the bulk densities in the absence of the colloid, the excess densities must be defined as the difference between the true hydrodynamic fields, $\bm{a}(\bm{r},t)$, and the bulk fields, $\bm{a}^+(\bm{r},t)$.  The validity of replacing the surface hydrodynamics by a boundary condition applied to the bulk hydrodynamic equations depends on the choice of $z_0$ since the required vanishing of the surface flow terms that appear in $\Sigma_a^s$ in Eqs.~(\ref{eq:gamma-sink}), (\ref{eq:massSink}) and (\ref{eq:momentumSink}) depends on the value of $z_0$.  For $z_0$ sufficiently far from the colloid surface, at least several times the interaction length $\ell_{\rm int}$ of the solvent-colloid interactions, the surface hydrodynamic equations should have a form analogous to the bulk equations, in which case the sink terms are negligible.  Numerical simulations of liquid systems near walls confirm that the thickness of the interface should be sufficient to establish interfacial hydrodynamic behavior\cite{CDDCE19}, and locate the interface at microscopic distances from the wall\cite{CWTS15}.

We have assumed that the fluid is unbounded and have not discussed the behavior of the hydrodynamic fields at the physical boundaries of the system. Although this may seem an oversimplification, in many experimental studies of isolated Janus motors in finite systems of macroscopic size\cite{PCYB10}, the motors remain far from the boundaries.  A motor of micron size has a typical active speed on the order of $3$ $\mu m/s$ and a long-time diffusion coefficient of $D_e\sim 2$ $\mu m^2/s$.  If a chemostat is applied to the system by feeding in reactants in a circular microfluidic chamber of diameter $650$ $\mu m$, as is done in Ref.~\onlinecite{PCYB10}, a motor initially at the center of the chamber will take on the order of $10^5$ seconds to diffuse to and interact with the system's outer boundaries.  For a motor near the feed source, the distance dependence of the motor to the boundary cannot be neglected, and the hydrodynamic fluid fields are no longer functions of their position relative to the colloid center alone. To describe the effect of external boundaries, a microscopic model for the interactions that confine or feed fluid particles into the system must be specified, which is important for channel geometries and systems maintained out of equilibrium by concentration gradients at the boundaries.  The resulting equations for the colloid and fluid densities will have a form similar to those presented in Sec.~\ref{sec:fluidHydro} and include additional terms that account for the coupling of the wall force to the fluid velocity.  Following the procedures outlined in Sec.~\ref{subsec:surf-layer}, hydrodynamic equations for dynamics of the fluid densities in the surface layer near the wall can be derived, and boundary conditions at the walls extracted from these expressions. 

The analysis presented here is easily generalized to non-isothermal systems where the internal energy density couples to the velocity and concentration fields, allowing for thermophoretic effects\cite{K83,SO93,SO97}.  The additional terms in the hydrodynamic equations arising from the internal energy density introduce an additional boundary condition\cite{K83,RKO77,GK19JSM} on the local inverse temperature $\beta (\bm{r},t) = (k_B T(\bm{r},t) )^{-1}$ at $z_0$.  Such considerations will be important if the catalytic conversion of reactants and products is either endothermic or exothermic or if external temperature gradients maintain the system out of equilibrium.

\section*{Acknowledgements}
Financial support from the National Sciences and Engineering Research Council of Canada is gratefully acknowledged. The authors thank Pierre Gaspard for many valuable discussions.

\appendix

\section{Interaction potentials and dynamics}\label{app:model}

The microscopic model is based on that introduced in the molecular derivation of the Langevin equation for an active colloid~\cite{RSGK20} where further details can be found. It is simplified to describe a Janus particle instead of a general active colloid.

The structureless solvent molecules have phase space coordinates, $\bm{x}^{N_S}=(\bm{x}_1, \bm{x}_2, \dots,\bm{x}_{N_s})=(\bm{r}^{N_S},\bm{p}^{N_S})$. The reactive species are made from $n_a$ chemically bound atoms with masses $\{m_k |  k=1,2,\dots, n_a\}$ and total mass $m$. The phase space coordinates of molecule $i$ are denoted by $\bm{x}_{i}^{n_a}=(\bm{x}_{(1)i},\bm{x}_{(2)i},\dots,\bm{x}_{(n_a)i}) =(\bm{r}_{i}^{n_a},\bm{p}_{i}^{n_a})$, while those for all $N_R$ reactive molecules are $\bm{x}_m^{N_R}=(\bm{r}_m^{N_R},\bm{p}_m^{N_R})$. For the entire fluid we have $\bm{x} = \{ \bm{x}^{N_{R}}_m, \bm{x}^{N_{S}} \}$.

The Hamiltonian for the system is
\begin{eqnarray}\label{eq:Ham_bath_col}
H&=&\frac{P^2}{2M}+K_{\rm rot} + \sum_{i=1}^N \Theta_i^S\frac{p_{i}^{2}}{2m}+\sum_{i=1}^N \Theta_i^R H_{mi} \nonumber\\
&&+U_{\rm f}(\bm{r}^{N_S},\bm{r}_m^{N_R})+U_{\rm I}(\bm{R},\bm{\theta},\bm{r}^{N_S},\bm{r}_m^{N_R}),\nonumber \\
&=& K+ H_0,
\end{eqnarray}
where the Hamiltonian for reactive molecule $i$ is
\begin{equation*}
%\label{eq:molecularH}
H_{mi}= \sum_{k=1}^{n_a} \frac{p_{(k)i}^2}{2m_k} + V_m(\bm{r}_i^{n_a}),
\end{equation*}
with $V_m(\bm{r}^{n_a}_i)$ the potential function for the chemically bonded nuclei in this molecule.  Interactions among fluid particles are denoted by $U_{\rm f}$, while $U_{\rm I}$ describes the interactions between the colloid and fluid particles. These interactions are assumed to have a short range so that interactions of the fluid molecules with the colloidal sites and those with each other are zero beyond a cut-off distance $\ell_{\rm int}$. In Eq.~(\ref{eq:Ham_bath_col}) the indicator functions $\Theta_i^{\nu}$, where $\Theta_i^{\nu}=1$ if molecule $i$ is species $\nu$ and $\Theta_i^{\nu}=0$ otherwise with $\nu \in \{S,R\}$, restrict the sums over fluid particles. The Hamiltonian for the fluid in the presence of a fixed colloid can be expressed in terms of the particle energies as
\begin{equation*}
%\label{eq:definition-ei}
H_0 = \sum_{i=1}^{N} \big( \Theta_i^S \frac{p_i^2}{2m} + \Theta_i^R H_{mi} + U_{{\rm f}i} + U_{{\rm I}i}\big)\equiv\sum_{i=1}^{N} e_i ,
\end{equation*}
where $U_{\rm f}=\sum_{i=1}^{N} U_{{\rm f}i}$ and $U_{\rm I}=\sum_{i=1}^{N} U_{{\rm I}i}=\sum_{\alpha =1}^{n_s} U_{\rm I}^\alpha$, and in the last equality we have written $U_{\rm I}$ as sum over sites.

The Liouville operator for the colloid $i{\cal L}_c$ is,
\begin{equation*} 
%\label{eq:Lc}
i{\cal L}_c = \frac{\bm{P}}{M} \cdot \bm{\nabla}_{\bm{R}} + \bm{F}  \cdot \bm{\nabla}_{\bm{P}} + \dot{\bm{\theta}} \cdot \bm{\nabla_\theta} + \bm{T} \cdot \bm{\nabla}_{\bm{L}} ,
\end{equation*}
where the force on the colloid is given by $\dot{\bm{P}}=\bm{F}= \sum_\alpha \bm{F}^\alpha$, where the contribution $\bm{F}^\alpha = -\bm{\nabla_R} U_{\rm I}^\alpha$ is due to fluid interactions $U_{\rm I}^\alpha$ with site $\alpha$, and  the torque on the colloid, is $\dot{\bm{L}}=\bm{T}=\sum_\alpha \bm{T}^\alpha$, where $\bm{T}^\alpha=\bm{S}^\alpha (\bm{R}) \wedge \bm{F}^\alpha$. Ref.~[\onlinecite{RSGK20}] gives additional information on the rigid body dynamics.

The Liouvillian for the fluid in the presence of the fixed colloid is,
\begin{eqnarray*}
%\label{eq:L0}
i\Liou_0 &=& \sum_{i=1}^N \Theta_i^S \Big(\frac{\bm{p}_i}{m} \cdot \bm{\nabla}_{\bm{r}_i} +\bm{F}_{i} \cdot \bm{\nabla}_{\bm{p}_i} \Big) \\
&&+\sum_{i=1}^N \Theta_i^R \sum_{k=1}^{n_a}\Big( \frac{\bm{p}_{(k)i}}{m_k} \cdot \bm{\nabla}_{\bm{r}_{(k)i}}  +\bm{F}_{(k)i} \cdot \bm{\nabla}_{\bm{p}_{(k)i}} \Big),\nonumber
\end{eqnarray*}
where $\bm{F}_{i} = -\bm{\nabla}_{\bm{r}_i} (U_{{\rm f} i}+U_{{\rm I} i})$ and $\bm{F}_{(k)i}=- \bm{\nabla}_{\bm{r}_{(k)i}} (U_{{\rm I} i} +V_{mi})$ are forces on solvent particle $i$ and atom $k$ in reactive molecule $i$, respectively. When an external force is present, the Liouvillian may be supplemented by an additional contribution, $i{\cal L}_{\rm{ext}} = \bm{F}_{\rm{ext}} \cdot \nabla_{\bm{P}}$.

\section{Fluxes of coarse-grained fields} \label{app:flux-coarse}

The expressions for the fluxes defined in Eq.~(\ref{eq:densityFlux}) are as follows: the fluxes of the species densities are
\begin{eqnarray*}
%\label{eq:rate-flux}
J_{R}(\bm{r})&=& \sum_{i=1}^N {\Theta}_i^{R} \dot{\xi}_i\delta(\xi_i-\xi^\ddag)\Delta(\bm{r}_{i}-\bm{r}), \\ \bm{j}_{\gamma}(\bm{r})&=&\sum_{i=1}^{N}\theta^\gamma_i(\xi_i) m^{-1}\bm{p}_{i}
 \Delta (\bm{r}_{i} -\bm{r}).
\end{eqnarray*}

The quantities that enter the fluxes of the other local densities are the fluid stress tensor, $\bm{\tau}(\bm{r})$
\begin{align}
\label{eq:stress-tensor}
\bm{\tau}&(\bm{r}) = \sum_{i=1}^N\Big[ \frac{\bm{p}_i\bm{p}_i}{m} -\frac{1}{2} \sum_{j\ne i}^N \bm{r}_{ij} \bm{\nabla}_{\bm{r}_i}U_{\rm f} \Big]\Delta(\bm{r}_{i}-\bm{r}),
\end{align}
written in the small gradient approximation where $\ell_{\rm{int}} / \ell_{\Delta} \ll 1$, and the local force on the fluid due to interactions with the colloid,
\begin{align*} 
%\label{eq:force-on-fluid}
\bm{F}_{\rm f}(\bm{r})&= -\sum_{i=1}^N \bm{\nabla}_{\bm{r}_i} U_{\rm I}
\Delta(\bm{r}_{i}-\bm{r}) =\sum_{\alpha} \bm{F}^\alpha_{\rm f}(\bm{r}).
\end{align*}
The total energy density $E(\bm{r})$ flux involves
\begin{eqnarray*}
%\label{eq:Eflux}
 J_E(\bm{r}) &=& (\bm{V}\cdot \bm{F}+ \bm{\omega} \cdot \bm{T})\delta(\bm{r}-\bm{R})\\
  &&+\sum_{\alpha} \bm{F}^\alpha_{\rm f}(\bm{r}) \cdot \big( \bm{V} + \bm{\omega} \wedge \bm{S}^\alpha (\bm{R}) \big),\nonumber \\
 \bm{j}_e(\bm{r}) &=& \bm{V} K \delta(\bm{r}-\bm{R})\\
 &&+\sum_{i=1}^N \Big[ \frac{\bm{p}_i}{m} e_i - \frac{1}{2m}
 \sum_{j \neq i} \bm{r}_{ji} \nabla_{\bm{r}_i} U_{\rm f} \cdot \bm{p}_i \Big]
 \Delta (\bm{r}_i-\bm{r}).\nonumber
\end{eqnarray*}
Here and elsewhere, we omit explicitly writing the energy density dependence on colloid variables for simplicity.

\section{Coarse-grained $\bm{\phi}$ fields} \label{app:coarse}

The conditions that lead to Eqs.~(\ref{eq:phi-fields}) are described below. To evaluate the term $\bm{A}(\bm{r}) \ast \bm{\phi}_A(\bm{r},t)$ that enters the local non-equilibrium density, it is sufficient to consider one contribution, say, $\bm{g}_N(\bm{r}) \ast  \bm{\phi}_v(\bm{r},t)= \bm{g}_N(\bm{r}) \ast \beta  \bm{v}(\bm{r},t)$. We then have
\begin{eqnarray*}
&&\bm{g}_N(\bm{r}) \ast \beta  \bm{v}(\bm{r},t)  = \sum_i \bm{p}_i \cdot \int d\bm{r} \; \beta \bm{v}(\bm{r},t) \Delta (\bm{r}_i - \bm{r}) \nonumber \\
&& \qquad \equiv \sum_i \bm{p}_i \cdot \beta\overline{\bm{v}} (\bm{r}_i,t)  .
\end{eqnarray*}
In the small gradient approximation where $\epsilon_\Delta = \ell_{\rm int}/\ell_{\Delta} \ll 1$, we note that since $\phi$ is a functional of the average coarse-grained densities $\bm{a}(\bm{r},t)$ , $\ell_{\Delta} \bm{\nabla}_{\bm{r}} \bm{\phi}(\bm{r},t) \sim \epsilon_\Delta$, and we can write
\begin{align*}
\overline{\bm{\phi}}&(\bm{r},t) = \int d\bm{r}^\prime \, \bm{\phi}(\bm{r}^\prime ,t) \Delta (\bm{r}-\bm{r}^\prime) \\
&= \int d \bm{r}^\prime \; \Delta (\bm{r}-\bm{r}^\prime) \Big[ \bm{\phi}(\bm{r},t) + (\bm{r}^\prime -\bm{r}) \cdot \bm{\nabla}_{\bm{r}} \bm{\phi}(\bm{r},t) \nonumber \\
& \qquad + \frac{1}{2} (\bm{r}^\prime -\bm{r}) (\bm{r}^\prime -\bm{r}) : \bm{\nabla}_{\bm{r}} \bm{\nabla}_{\bm{r}} \bm{\phi}_(\bm{r},t) + \dots \Big] .
\end{align*}
For a spherically symmetric coarse-graining function, such as a normal distribution with zero mean and covariance $\ell_{\Delta}^2 \bm{1}$, we find the coarse-grained fields $\tilde{\phi}$ coincide with the fields $\phi$ to second order in $\epsilon_\Delta$,
\begin{align*}
\overline{\bm{\phi}}(\bm{r},t) &= {\bm{\phi}}(\bm{r},t) + \frac{1}{2} \ell_{\Delta}^2 \bm{\nabla}_{\bm{r}}^2 \bm{\phi}(\bm{r}^,t) = {\bm{\phi}}(\bm{r},t) + O(\epsilon_\Delta^2).
\end{align*}
It follows that
\begin{equation*}
\langle \bm{p}_i \rangle_t = m \overline{ \bm{v}}(\bm{r_i},t) = m \bm{v}(\bm{r}_i,t) + O(\epsilon_\Delta^2),
\end{equation*}
and hence, by Taylor expansion, we find
\begin{eqnarray*}
&&\langle \bm{g}_N (\bm{r}) \rangle_t = m \Big\langle \sum_i \bm{v} (\bm{r}_i,t) \Delta (\bm{r} - \bm{r}_i) \Big\rangle_t \\
 && \qquad = m \bm{v}(\bm{r},t) n(\bm{r},t)
 + \bm{\nabla}_{\bm{r}} \bm{v}(\bm{r},t) \cdot \bm{n}^{(1)}(\bm{r},t) ,\nonumber
\end{eqnarray*}
where
\begin{align*}
\bm{n}^{(1)}(\bm{r},t) &= \Big\langle \sum_i \big( \bm{r}_i - \bm{r} \big) \Delta (\bm{r} - \bm{r}_i) \Big\rangle_t \\
&=-\ell_{\Delta}^2 \bm{\nabla}_{\bm{r}} n(\bm{r},t) \sim O(\epsilon_\Delta).
\end{align*}
Thus,  we have $\langle \bm{g}_N (\bm{r}) \rangle_t=m \bm{v}(\bm{r},t) n(\bm{r},t) + O(\epsilon_\Delta^2)$.

More generally, one can show that $\overline{\bm{\phi}}(\bm{r},t) = {\bm{\phi}}(\bm{r},t) + O(\epsilon_\Delta^2)$; thus, to terms of order $\epsilon_\Delta^2$ we obtain Eqs.~(\ref{eq:phi-fields}) in the text.

\section{Slip velocity} \label{app:slip}

We show that the last term in Eq.~(\ref{eq:rhoc-term}) involving $\bm{F}_{{\rm f},t}^{\alpha} (\bm{r}^\prime)$,
\begin{equation*}
%\label{eq:Z}
\mathcal{Z}=\sum_\alpha \bm{F}_{{\rm f},t}^\alpha(\bm{r}^\prime) \ast  \big(\bm{v}(\bm{r}^\prime,t) - \bm{V}_c - \bm{\omega}_c \wedge \bm{S}^{\alpha}(\bm{R})\big),
\end{equation*}
can be expressed in terms of the slip velocity.
The force $\bm{F}_{{\rm f},t}^\alpha$ can be written as the sum of contributions from each fluid particle $i$,
$\bm{F}_{{\rm f},t}^{\alpha}=\sum_{i=1}^N  \bm{F}_{{\rm f}i,t}^\alpha(\bm{r}^\prime)$.
\begin{figure}[htbp]
\centering
\resizebox{0.8\columnwidth}{!}{
      \includegraphics{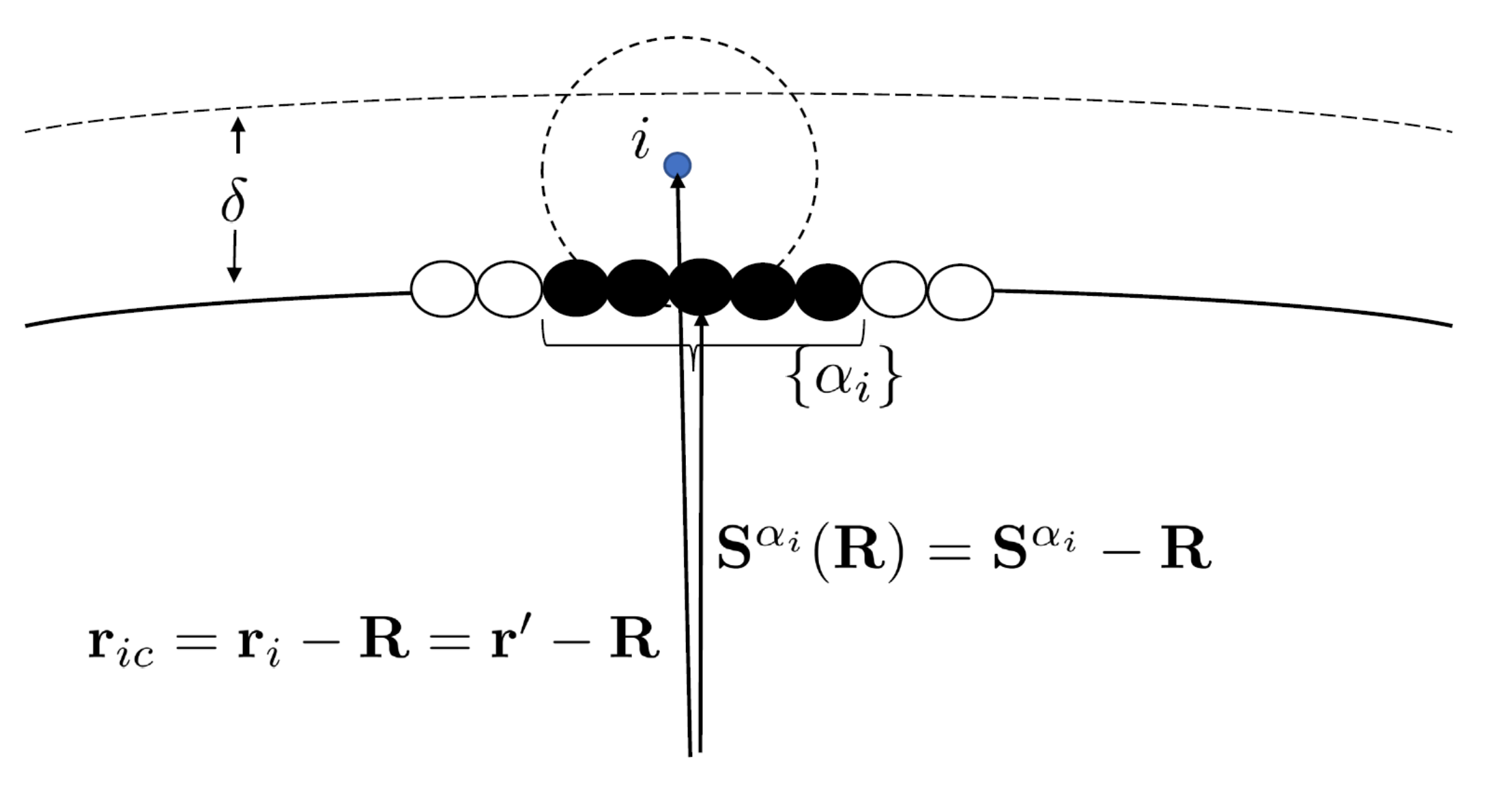}}
        \caption{ Schematic diagram of a small portion of the surface of a colloid at position $\bm{R}$ with a large radius showing a fluid particle $i$ at position $\bm{r}_{i}$. corresponding to the field point $\bm{r}^\prime$.  This particle $i$ at $\bm{r}_{ic}=\bm{r}_{i}-\bm{R}$ interacts with the set $\{\alpha_i\}$ of black sites centered at site $\alpha_i$ located $\bm{S}^{\alpha_i}(\bm{R})$ relative to the center of the colloid. The dashed line at a distance $\delta$ from the surface sites denotes the region where the interaction potential is non-zero. The dashed circle around the particle $i$ denotes the coarse-graining domain.
    } \label{fig:alpha}
\end{figure}
Figure~\ref{fig:alpha} shows a small portion of the surface of a large colloid with radius $R_J$. In this figure, we see that particle $i$ at position $\bm{r}_i$ at field point $\bm{r}'$ interacts with a small set of surface sites $\{\alpha_i\}$ centered at site $\alpha_i$ closest to it. Therefore, the sum of $\bm{F}_{{\rm f}i,t}^\alpha$ on $\alpha$ can be restricted to sites in this set; hence,
\begin{equation*}
\mathcal{Z} \approx \sum_{i=1}^N\sum_{\alpha  \in \{ \alpha_i \}} \bm{F}_{{\rm f}i,t}^\alpha(\bm{r}^\prime) \ast \big(\bm{v}(\bm{r}^\prime,t) - \bm{V}_c - \bm{\omega}_c \wedge \bm{S}^{\alpha}(\bm{R})\big).
\end{equation*}
Since the set $\{ \alpha_i \}$ subtends a small surface angle we may take $\bm{S}^{\alpha}(\bm{R}) \approx \bm{S}^{\alpha_i}(\bm{R}) = \bm{r}^\prime-\bm{R}+ O(\delta/R_J)$ where $\delta \sim \ell_{\rm int}$ is the thickness of the boundary layer. We observe that $ \sum_{i=1}^N \Big(\sum_{\alpha  \in \{ \alpha_i \}}\bm{F}_{{\rm f}i,t}^\alpha \Big) \Delta (\bm{r}_i - \bm{r}^\prime) \approx {\bm{F}}_{{\rm f},t}(\bm{r}^\prime)$ since the coarse-graining length is larger than the interaction range of the potentials.  We obtain
\begin{eqnarray*}
\mathcal{Z} &\approx&  {\bm{F}}_{{\rm f},t}(\bm{r}^\prime) \ast \big(\bm{v}(\bm{r}^\prime,t) - \bm{V}_c - \bm{\omega} \wedge (\bm{r}^\prime-\bm{R})\big)\nonumber \\
&\equiv& {\bm{F}}_{{\rm f},t}(\bm{r}^\prime) \ast  \bm{v}_{\rm sl}(\bm{r}^\prime,t)),
\end{eqnarray*}
with $\mid\bm{r}^\prime-\bm{R}\mid=R_J$.

\section{Fluxes that enter hydrodynamic equations} \label{app:hydro-fluxes}

Writing Eq.~(\ref{eq:a-hydro}), we have
\begin{eqnarray}\label{eq:formal-hydro}
\partial_t \bm{a}(\bm{r},t) &=& \bm{\mathcal{E}}_a(\bm{r},t)-\bm{L}^{(0 0)}_{a a'}(\bm{r},\bm{r}',t)\ast \bm{\phi}_{a'}(\bm{r}',t)\nonumber \\
 &-&  \bm{L}^{(0 1)}_{a a'}(\bm{r},\bm{r}',t)\ast \bm{\nabla}_{r'} \bm{\phi}_{a'}(\bm{r}',t)\nonumber\\
 &+&   \bm{\nabla{}}_r \cdot \bm{L}^{(1 0)}_{a a'}(\bm{r},\bm{r}',t)\ast \bm{\phi}_{a'}(\bm{r}',t)\nonumber \\
 &+&   \bm{\nabla{}}_r \cdot \bm{L}^{(1 1)}_{a a'}(\bm{r},\bm{r}',t)\ast \bm{\nabla}_{r'} \bm{\phi}_{a'}(\bm{r}',t).
\end{eqnarray}
The expressions for the $\bm{\phi}_a$ fields are given in Eq.~(\ref{eq:phi-fields}), with the replacement of the velocity field by the slip velocity field. Using these results we can now write explicit expressions for $\bm{L}_{a a'}^{(\ell 0)} \ast \phi_{a'}$ and $\bm{L}_{a a'}^{(\ell 1)} \ast \bm{\nabla}_{\bm{r}'} \phi_{a'}$ terms in Eq.~(\ref{eq:formal-hydro}) that were used to obtain the hydrodynamic equations~(\ref{eq:solute})-(\ref{eq:momentum}):
\begin{align*}
\bm{L}_{a a'}^{(\ell 0)} \ast \phi_{a'} &= \beta \bm{L}^{(\ell 0)}_{a v} \ast  \bm{v}_{\rm sl}
+ \beta \bm{L}_{a R}^{(\ell 0)} \ast  {\mathcal{A}} \\
\bm{L}_{a a'}^{(\ell 1)} \ast \bm{\nabla}_{\bm{r}'} \phi_{a'} &= \beta \bm{L}_{a v}^{(\ell 1)} \ast  \bm{\nabla}_{\bm{r}'}\bm{v} +\beta \bm{L}^{(\ell 1)}_{a\gamma} \ast \bm{\nabla}_{\bm{r}'} \widetilde{\mu}_{\gamma},\nonumber
\end{align*}
where we have not displayed the dependence of the transport coefficients and thermodynamic fields on $\bm{r}'$ and $t$, and
\begin{eqnarray*}
\beta \bm{L}^{(\ell 0)}_{a v} &=&  \int_{0}^\infty d\tau \; \left\langle \bm{j}^{(\ell)}_{a,t}(\bm{r},\tau) \bm{F}_{{\rm{f}},t}(\bm{r}') \right\rangle_t  \\
\beta \bm{L}^{(\ell 1)}_{a v} &=&  \int_{0}^\infty d\tau \; \left\langle \bm{j}^{(\ell)}_{a,t}(\bm{r},\tau) \bm{\tau}_{{\rm{f}},t}(\bm{r}') \right\rangle_t \nonumber \\
\beta \bm{L}^{(\ell 0)}_{a R} &=& \int_{0}^\infty d\tau \; \left\langle \bm{j}^{(\ell)}_{a,t}(\bm{r},\tau) \bm{J}_{R,t} (\bm{r}') \right\rangle_t  \nonumber \\
\beta \bm{L}^{(\ell 1)}_{a \gamma} &=& \int_{0}^\infty d\tau \; \left\langle \bm{j}^{(\ell)}_{a,t}(\bm{r},\tau) \bm{j}_{\gamma,t} (\bm{r}') \right\rangle_t . \nonumber
\end{eqnarray*}

The Euler terms can be computed from the local equilibrium average of the fluxes, and to order $\epsilon_\Delta^2$ are
\begin{eqnarray}
\mathcal{E}_{\gamma} (\bm{r},t) &=& -\bm{\nabla}_r\cdot
\big( n_\gamma (\bm{r},t) \bm{v}(\bm{r},t) \big) -\nu_\gamma \langle J_R(\bm{r}) \rangle_t \nonumber \\
\mathcal{E}_n (\bm{r},t) &=&-\bm{\nabla{}}_r\cdot{}
\big(n(\bm{r},t) \bm{v}(\bm{r},t\big) \nonumber \\
\mathcal{E}_g(\bm{r},t) &=&
-\bm{\nabla}_{\bm{r}} \cdot \left\langle \bm{\tau} (\bm{r}) \right\rangle_t
+\langle \bm{F}_{\rm f}(\bm{r})\rangle_{t} \nonumber \\
&=& -\bm{\nabla{}}_r\cdot{}\big(mn(\bm{r},t)\bm{v}(\bm{r},t)\bm{v}(\bm{r},t)\big)\nonumber \\
 &&\qquad -\bm{\nabla{}}_r\cdot{}\langle{}\bm{\tau{}}^\ddag(\bm{r})\rangle_{t}
+\langle{}\bm{F}_{\rm f}(\bm{r})\rangle_{t} \label{eq:euler}
%\mathcal{E}_e (\bm{r},t) &=&
%-\bm{\nabla}_{\bm{r}} \cdot \left\langle \bm{j}_e (\bm{r}) \right\rangle_t \label{eq:euler-energy} \\
%&&\quad
%+ \sum_\alpha \langle \bm{F}^\alpha_{\rm f}(\bm{r})\rangle_{t} \cdot \left( \bm{V} +\bm{\omega} \wedge %\bm{S}^\alpha (\bm{R}) \right) \nonumber \\
%&=& -\bm{\nabla}_{\bm{r}} \cdot \Big( \bm{v}(\bm{r},t) \big[ e^\ddag(\bm{r},t) + \langle \bm{\tau}^\ddag %\rangle_t \nonumber \\
%&& \qquad + \frac{1}{2}m n(\bm{r},t) v^{2}(\bm{r},t) \big] \Big) \nonumber \\
%&&\quad
%+ \sum_\alpha \langle \bm{F}^\alpha_{\rm f}(\bm{r})\rangle_{t} \cdot \left( \bm{V} +\bm{\omega} \wedge %\bm{S}^\alpha (\bm{R}) \right) . \nonumber
\end{eqnarray}
The stress tensor $\bm{\tau{}}^{\ddag}(\bm{r})$ has an expression similar to that in Eq.~(\ref{eq:stress-tensor}) with $\bm{p}_{i}$ replaced by $\bm{p}_{i}^{\ddag}=\bm{p}_{i}-m\bm{v}(\bm{r}_i, t)$. In Appendix~\ref{app:react-flux}, we show that  $\langle J_R(\bm{r}) \rangle_t$ is a reactive contribution proportional to the fluid velocity and may be neglected in most circumstances.

%Using the definition of the internal energy density and Eqs.~(\ref{eq:euler-momentum}) and %(\ref{eq:euler-energy}), the Euler terms for the internal energy density $e^+ (\bm{r},t)$ are
%\begin{align}
%\mathcal{E}_{e^\ddag} (\bm{r},t) &= -\bm{\nabla}_{\bm{r}} \cdot \Big[ \bm{v}(\bm{r},t) e^+(\bm{r},t) + %\bm{v}(\bm{r},t)\cdot \langle \bm{\tau}^\ddag \rangle_t\Big] \nonumber \\
%&\qquad - \langle \bm{F}_{\rm f}(\bm{r})\rangle_{t} \cdot {\bm{v}}_s (\bm{r},t) ,
%\end{align}
%where we have used Eq.~(\ref{eq:Falpha-approx}) to write $\langle \bm{F}_{\rm f}(\bm{r})\rangle_{t} \cdot %{\bm{v}}_s (\bm{r},t) \approx \sum_\alpha \langle \bm{F}^\alpha_{\rm f}(\bm{r})\rangle_{t} \cdot %{\bm{v}}^\alpha (\bm{r},t)$

The full expressions for the components of the $\bm{\mathcal{J}}^{(0)}_A(\bm{r},t)$ and $\bm{\mathcal{J}}^{(1)}_A(\bm{r},t)$ fluxes are:
\begin{align*}
{\bm{\mathcal{J}}}^{(0)}_\gamma &= -\beta {L}^{(00)}_{\gamma R} \ast {\mathcal{A}}
- \beta {L}^{(00)}_{\gamma v} \ast \bm{v}_{\rm sl} \nonumber \\
& \quad - \beta \bm{L}^{(01)}_{\gamma \gamma'} \ast \bm{\nabla{}}_{r'} \widetilde{\mu}_{\gamma'} - \beta \bm{L}^{(01)}_{\gamma v} \ast  \bm{\nabla{}}_{r'}\bm{v} \\
%\label{eq:tilde-Jgamma-zero} \\
{\bm{\mathcal{J}}}^{(0)}_\rho &= 0 \\
%\label{eq:jn-zero} \\
{\bm{\mathcal{J}}}^{(0)}_{v} &= -\beta {L}^{(00)}_{v R} \ast  {\mathcal{A}}
- \beta {L}^{(00)}_{v v}  \ast  \bm{v}_{\rm sl} \nonumber \\
& \quad -\beta \bm{L}^{(01)}_{v \gamma'} \ast \bm{\nabla{}}_{r'} \widetilde{\mu}_{\gamma'}
-\beta \bm{L}^{(01)}_{v v} \ast \bm{\nabla}_{r'} \bm{v} . 
%\label{eq:jv-zero}
%&&\bm{\mathcal{J}}^{(0)}_{e^\ddag} =  - \bm{v}_s \cdot {\mathcal{J}}_{v}^{(0)} \label{eq:jep-zero}.
\end{align*}
and
\begin{align*}
{\bm{\mathcal{J}}}^{(1)}_\gamma &= -\beta {L}^{(10)}_{\gamma R} \ast {\mathcal{A}}
        - \beta {L}^{(10)}_{\gamma v} \ast {\bm{v}_{\rm sl}}  \nonumber \\
& \quad -\beta \bm{L}^{(11)}_{\gamma \gamma'} \ast \bm{\nabla{}}_{r'} \widetilde{\mu}_{\gamma'} -\beta \bm{L}^{(11)}_{\gamma v} \ast \bm{\nabla{}}_{r'}\bm{v}  \\
% \label{eq:jgamma-one}\\
{\bm{\mathcal{J}}}^{(1)}_\rho &=0 \\
{\bm{\mathcal{J}}}^{(1)}_{v} &=  -\beta \bm{L}^{(10)}_{v R} \ast  {\mathcal{A}}  - \beta \bm{L}^{(10)}_{v v} \ast {\bm{v}_{\rm sl}} \nonumber \\
& \quad -\beta \bm{L}^{(11)}_{v \gamma'} \ast \bm{\nabla{}}_{r'} \widetilde{\mu}_{\gamma'}  - \beta \bm{L}^{(11)}_{v v,t} \ast  \bm{\nabla{}}_{r'}\bm{v}  
%\label{eq:jv-one}
%&&{\bm{\mathcal{J}}}^{(1)}_{e^\ddag} = \bm{L}^{(10)}_{e^\ddag R} \ast {\mathcal{A}} + \bm{L}^{(10)}_{e^\ddag v} \ast \beta {\bm{v}_s} + \bm{L}^{(11)}_{e^\ddag e^+\ddag} \cdot \bm{\nabla}_{\bm{r}^\prime} \beta\nonumber \\
%&&\qquad -\bm{L}^{(11)}_{e^\ddag \gamma'} \ast \bm{\nabla{}}_{r'}\left( \beta \widetilde{\mu}_{\gamma'} \right)
%- \bm{L}^{(11)}_{e^\ddag v} \ast \beta \bm{\nabla{}}_{r'}\bm{v},
%\label{eq:jep-one}
\end{align*}
where the spatial and time dependence of all quantities has been omitted for notational simplicity.

The full expressions for the fluid hydrodynamic equations in the presence of a moving active Janus colloid given in Eqs.~(\ref{eq:solute})-(\ref{eq:momentum}) follow from these results.

\section{Surface equations and fluxes} \label{app:surf-flux}

We now use several physically motivated approximations to express the fluxes in Eq.~(\ref{eq:a-surface-int}) in a more explicit and tractable form. The $\bm{\mathcal{J}}^{(0)}_{a0}(\bm{r}_\parallel,t)$ fluxes exist only in the boundary layer and can be written as 
\begin{eqnarray*}
%\label{eq:J(0)-fluxes-local}
&&\bm{\mathcal{J}}^{(0)}_{a0}(\bm{r}_\parallel,t)= -\int dz \;\Big[ \bm{L}^{(0 0)}_{a a'}(\bm{r},\bm{r}',t)\ast \bm{\phi}_{a'}(\bm{r}',t) \nonumber \\
&& \qquad +  \bm{L}^{(0 1)}_{a a'}(\bm{r},\bm{r}',t)\ast \bm{\nabla}_{r'} \bm{\phi}_{a'}(\bm{r}',t)\Big] \nonumber \\
&& \qquad \qquad \quad = -\int dz \;\Big[ \bm{L}^{(0 0)}_{a a'}(\bm{r},t)\cdot \bm{\phi}_{a'}(\bm{r},t) \nonumber \\
&& \qquad +  \bm{L}^{(0 1)}_{a a'}(\bm{r},t)\cdot \bm{\nabla}_{r} \bm{\phi}_{a'}(\bm{r},t)\Big].
\end{eqnarray*}
The second equality follows since the correlation functions in $\bm{L}_{aa'}^{(0 \ell')}(\bm{r},\bm{r}',t)$ vanish for $|\bm{r}-\bm{r}'| > \xi_{\text{mic}}$; thus, $\bm{r}'$ is also restricted to lie in this spatial region. We then use the fact that the conjugate $\bm{\phi}_a(\bm{r}',t)$ fields are smoothly varying spatial functions, so they and their derivatives may be expanded in a Taylor series around the point $\bm{r}$. Consequently,
\begin{align*}
\bm{j}_{a',t}^{(0)}(\bm{r}') \ast \bm{\phi}_{a'}(\bm{r}',t) &= \bm{j}_{a',t}^{(0)} \cdot \bm{\phi}_{a'}(\bm{r},t) + \dots \\
\bm{j}_{a',t}^{(1)}(\bm{r}') \ast \bm{\nabla}_{\bm{r}'} \bm{\phi}_{a'}(\bm{r}',t) &= \bm{j}_{a',t}^{(1)} \cdot \bm{\nabla}_{\bm{r}} \bm{\phi}_{a'}(\bm{r},t) + \dots ,
\end{align*}
where the $\bm{j}^{(\ell)}_{a,t} = \int d\bm{r}' \, \bm{j}^{(\ell)}_{a,t}(\bm{r}')$ are the integrated microscopic fluxes.  Finally, since $\bm{r}$ is confined to the boundary layer, and again using smoothness of the coarse-grained $\phi_{a'}(\bm{r})$ fields, we can expand them in a Taylor series about $z_0$,
\begin{equation*}
\bm{\phi}_{a'}(z,\bm{r}_\parallel) = \bm{\phi}_{a'}(z_0,\bm{r}_\parallel) + (z-z_0) \partial_z  \bm{\phi}_{a'}(z,\bm{r}_\parallel)|_{z_0} + \cdots
\end{equation*}
Taking $z_0$ to be at the outer part of the boundary layer where the $\bm{\phi}_{a}$ fields closely approximate their bulk values, $\bm{\phi}_{a}^+$, we obtain
\begin{eqnarray*}
%\label{eq:J0-fluxes-local2}
\bm{\mathcal{J}}^{(0)}_a(\bm{r},t)&\approx& -\Big[ \bm{L}^{(0 0)}_{a a'}(\bm{r}_\parallel,t)\cdot \bm{\phi}_{a'}^+(z_0,\bm{r}_\parallel,t)\nonumber \\
&& + \bm{L}^{(0 1)}_{a a'}(\bm{r}_\parallel,t)\cdot \bm{\nabla}_{r_\parallel} \bm{\phi}_{a'}^+(z_0,\bm{r}_\parallel,t)\Big],
\end{eqnarray*}
with transport coefficients
\begin{eqnarray*}
\bm{L}^{(\ell \ell')}_{a a'}(\bm{r}_\parallel,t)&=&\int_0^\infty d\tau \; \langle \bm{j}^{(\ell)}_{a,t}(\bm{r}_\parallel,\tau) \bm{j}^{(\ell')}_{a',t}\rangle_{t}, \label{eq:Ljj-local0}
\end{eqnarray*}
where $\bm{j}^{(\ell)}_{a,t}(\bm{r}_\parallel,\tau)=\int dz \; \bm{j}^{(\ell)}_{a,t}(z,\bm{r}_\parallel,\tau)$.

Using Eqs.~(\ref{eq:J-fluxes}) and (\ref{eq:Jtilde-full}) the ${\bm{\mathcal J}}^{(1)s}_{a0} (z,\bm{r}_\parallel ,t)$ fluxes are given by
\begin{eqnarray}
\label{eq:J1-fluxes-local}
&&\bm{\mathcal{J}}^{(1)s}_{a0}(\bm{r}_\parallel,t)= -\int dz \;\Big[ \bm{L}^{(1 0)}_{a a'}(\bm{r},\bm{r}',t)\ast \bm{\phi}_{a'}(\bm{r}',t) \nonumber \\
&& \qquad +  \bm{L}^{(1 1)}_{a a'}(\bm{r},\bm{r}',t)\ast \bm{\nabla}_{r'} \bm{\phi}_{a'}(\bm{r}',t) \nonumber \\
&& \qquad - \theta(z-z_0)  \bm{L}^{+}_{a a'}(\bm{r},\bm{r}',t)\ast \bm{\nabla}_{r'} \bm{\phi}_{a'}^+(\bm{r}',t)\Big].
\end{eqnarray}
The first term in this equation can be evaluated as above to give
\begin{eqnarray*}
&&\int dz \; \bm{L}^{(1 0)}_{a a'}(\bm{r},\bm{r}',t)\ast \bm{\phi}_{a'}(\bm{r}',t) \nonumber \\
&& \qquad \qquad \approx \bm{L}^{1 0)}_{a a'}(\bm{r}_\parallel,t)\cdot \bm{\phi}_{a'}^+(z_0,\bm{r}_\parallel,t),  
\end{eqnarray*}
while the second term requires further analysis since there are discontinuities in $z$ that prevent straightforward localization. 

For $z \gg z_0$, the transport coefficient $\bm{L}^{(11)}_{a a'}(z,\bm{r}_\parallel,\bm{r}', t)$ that appears in the surface flux is equal to its value in the bulk, $\bm{L}^{+}_{a a'}(\bm{r}, \bm{r}',t)$, and hence this contribution to the surface flux $\bm{\mathcal{J}}^{(1)s}_{a}(\bm{r},t)$ is non-zero only in the boundary layer. To evaluate it, we write the $\bm{\phi}_a$ fields in terms of their surface and bulk contributions, $\bm{\phi}_{a}(\bm{r},t)=\bm{\phi}_{a}^s(\bm{r},t) +\theta(z-z_0) \bm{\phi}_{a}^+(\bm{r},t)$, 
so that $\bm{\nabla}_r \bm{\phi}_{a}(\bm{r},t) =\bm{\nabla}_r \bm{\phi}_{a}^s(\bm{r},t)+ \hat{\bm{z}} \delta(z-z_0) \bm{\phi}_{a}^+(\bm{r},t) +\theta(z-z_0) \bm{\nabla}_r \bm{\phi}_{a}^+(\bm{r},t)$. Next, we evaluate the surface terms in the lowest order multipole approximation, $\bm{\phi}_{a}^s(\bm{r},t) \approx \delta (z-z_0)\bm{\phi}_{a0}^s(\bm{r}_\parallel,t)$. After localizing in $\bm{r}_\parallel$, then have
\begin{eqnarray}
\label{eq:J1-fluxes-local3}
&&\int dz \;\Big[ \bm{L}^{(1 1)}_{a a'}(\bm{r},\bm{r}',t)\ast \bm{\nabla}_{r'} \bm{\phi}_{a'}(\bm{r}',t)  \nonumber \\
&& \qquad - \theta(z-z_0)  \bm{L}^{+}_{a a'}(\bm{r},\bm{r}',t)\ast \bm{\nabla}_{r'} \bm{\phi}_{a'}^+(\bm{r}',t)\Big] \nonumber \\
&& \approx \; \bm{L}^{(1 1)}_{a a'}(\bm{r}_\parallel,z_0,t)\cdot \big(\bm{\nabla}_{r_\parallel} \bm{\phi}_{a'0}(\bm{r}'_\parallel,t)+\hat{\bm{z}} \bm{\phi}^+_{a'}(z_0,\bm{r}_\parallel,t)\big) \nonumber \\
&& \quad +  \bm{L}^{\theta}_{a a'}(\bm{r}_\parallel,t)\cdot \bm{\nabla}_{r_\parallel} \bm{\phi}_{a'}^+(z_0,\bm{r}_\parallel,t),
\end{eqnarray}
with 
\begin{eqnarray*}
\bm{L}^{\theta}_{a a'}(\bm{r}_\parallel,t)&=& \int dz' \; \big[\bm{L}^{(1 1)}_{a a'}(\bm{r}_\parallel,z',t)\theta(z'-z_0) \nonumber \\
&&  -\theta(z-z_0) \bm{L}^{+}_{a a'}(\bm{r}_\parallel,z',t) \big].\nonumber    
\end{eqnarray*}
and 
\begin{eqnarray*}
\bm{L}^{(\ell \ell')}_{a a'}(\bm{r}_\parallel,z_0 ,t)&=&\int_0^\infty d\tau \; \langle \bm{j}^{(\ell)}_{a,t}(\bm{r}_\parallel,\tau) \bm{j}^{(\ell')}_{a',t}(z_0)\rangle_{t}. \nonumber
\end{eqnarray*}
In writing Eq.~(\ref{eq:J1-fluxes-local3}), we dropped higher order terms involving $z$ derivatives of the fluxes. With these results, we can write
\begin{eqnarray*}
\label{eq:J1-fluxes-local-last}
&&\bm{\mathcal{J}}^{(1)s}_{a0}(\bm{r}_\parallel,t) \approx 
\bm{L}^{1 0)}_{a a'}(\bm{r}_\parallel,t)\cdot \bm{\phi}_{a'}^+(z_0,\bm{r}_\parallel,t)  \nonumber \\
&& \quad +  \bm{L}^{(1 1)}_{a a'}(\bm{r}_\parallel,z_0,t)\cdot \big(\bm{\nabla}_{r_\parallel} \bm{\phi}^s_{a'0}(\bm{r}'_\parallel,t)+\hat{\bm{z}} \bm{\phi}^+_{a'}(z_0,\bm{r}_\parallel,t)\big) \nonumber \\
&& \quad +  \bm{L}^{\theta}_{a a'}(\bm{r}_\parallel,t)\cdot \bm{\nabla}_{r_\parallel} \bm{\phi}_{a'}^+(z_0,\bm{r}_\parallel,t).
\end{eqnarray*}

The final simplification is the evaluation of the correlation functions in the homogeneous ensemble in the parallel direction. The Janus colloid surface has two distinct catalytic and noncatalytic hemispheres, which suggests that the use of the homogeneous ensemble should be carried out separately on the two hemispheres. While this can be done, it will lead to local surface transport coefficients that depend on the hemisphere. This may be necessary in some instances but is not usually implemented in standard boundary conditions. To this end, we assume that the interactions of the fluid species with the colloid beads do not depend on the bead identity so that the parallel homogeneous ensemble can be assumed for the entire colloid surface. The exceptions are the reactive terms since chemical reaction occurs only on the catalytic beads. Even in this case, we can suppose that the entire colloid is catalytic to obtain reaction rate coefficients per unit surface area and then restrict the reaction rate to the catalytic hemisphere.  

Taking this approach, the corresponding local equilibrium density is analogous to that in Eq.~(\ref{eq:local_eq-homo}) and is given by
 \begin{equation*}
%\label{eq:local_eq-homo-parallel}
\rho_{H,\parallel}(t) =\frac{\Pi_\lambda (N_\lambda ! h^{3N_\lambda})^{-1}e^{\bm{A} \cdot \bm{\phi}_{A}(z_0,\bm{r}_\parallel,t)}}{Tr[
\Pi_\lambda (N_\lambda ! h^{3N_\lambda})^{-1}e^{\bm{A} \cdot \bm{\phi}_{A}(z_0,\bm{r}_\parallel,t)}]}.
\end{equation*}
The resulting expressions for the $\bm{L}^{(\ell \ell')s}_{a a',H}$ surface coefficients, are
\begin{align*}
%\label{eq:L-Hp}
\bm{L}^{(\ell \ell')}_{a a',H}(\bm{r}_\parallel,t) &=\frac{1}{A}\int_0^\infty d\tau \; \left\langle \bm{j}^{(\ell)}_{a,t}(\tau) \bm{j}^{(\ell')}_{a',t} \right\rangle_{H}(\bm{r}_\parallel ,t) 
%\\ \bm{L}^{(11)}_{a a',H}(z_0,\bm{r}_\parallel,t) &=\frac{1}{A}\int_0^\infty d\tau \; \left\langle \bm{j}^{(1)}_{a,t}(\tau) \bm{j}^{(1)}_{a',t} (z_0)\right\rangle_{H}(\bm{r}_\parallel ,t), \nonumber
\end{align*}
with similar expressions for other correlation functions.

The analysis above can be extended beyond the zeroth-order multipoles of the surface excess densities.~\cite{RBO78} However, including higher-order multipoles that couple to the dynamics of the bulk hydrodynamic equations through boundary conditions leads to a corresponding increase in the number of surface transport coefficients about which little is known.

\section{Reactive flux $\langle J_R(\bm{r}) \rangle_t$}\label{app:react-flux}

The reactive contribution $\langle J_R(\bm{r}) \rangle_t$ in the first equation in (\ref{eq:euler}) can be evaluated as follows: the time derivative of the species variable for particle $i$ is
\begin{eqnarray*}
&&\dot{\theta}_i^\gamma(\xi_i(\bm{r}_i^{n_a})) =
%\Theta_i^R\sum_{k=1}^{n_a} \frac{\bm{p}_{(k)i}}{m_k} \cdot \bm{\nabla}_{r_{(k)i}}H_\gamma(\xi_i(\bm{r}_i^{n_a}))\nonumber \\
%&=&
\Theta_i^R\sum_{k=1}^{n_a} \frac{\bm{p}^\ddagger_{(k)i}}{m_k} \cdot \bm{\nabla}_{r_{(k)i}}H_\gamma(\xi_i(\bm{r}_i^{n_a}))\nonumber \\
&& \qquad+\Theta_i^R \bm{v}(\bm{r},t) \cdot \sum_{k=1}^{n_a}\bm{\nabla}_{r_{(k)i}}H_\gamma(\xi_i(\bm{r}_i^{n_a})).
\end{eqnarray*}
On average, the first term vanishes so that the reactive flux is given by
\begin{eqnarray*}
%\label{eq:average-R-flux}
&&\nu_\gamma \langle J_R(\bm{r}) \rangle_t= \\
&&\Big\langle \sum_{i=1}^N \Theta_i^R \sum_{k=1}^{n_a}\bm{\nabla}_{r_{(k)i}}H_\gamma(\xi_i(\bm{r}_i^{n_a}))\Delta(\bm{r}_{ic}-\bm{r})\Big\rangle_t \cdot \bm{v}(\bm{r} , t).\nonumber
\end{eqnarray*}
This contribution is proportional to the fluid velocity field $\bm{v}(\bm{r},t)$. Its prefactor is nonzero only at the barrier top and is a fluid-velocity-dependent term in the surface layer.

We now consider the computation of the dissipative reactive flux contribution to $\bm{\mathcal{J}}^{(0)}_{\gamma 0}(\bm{r}_\parallel ,t)$ in Eq.~(\ref{eq:J0gammasH-2}).
The projected reactive flux that enters the reactive flux correlation function is
\begin{eqnarray}\label{eq:proj-J-R}
&&J^D_{R,t}(\bm{r})= \mathcal{Q}_{A}(t) J_R(\bm{r})= J_R(\bm{r}) -\langle J_R(\bm{r}) \rangle_t
 \\
&&\qquad -\langle J_R(\bm{r}) \widetilde{\bm{A}}(\bm{r}_{1})\rangle_t\ast{}\langle{}\widetilde{\bm{A}}\widetilde{\bm{A}}\rangle_{t}^{-1}(\bm{r}_{1},\bm{r}_{2})
\ast{}\widetilde{\bm{A}}(\bm{r}_{2}) ,\nonumber
\end{eqnarray}
where $\tilde{\bm{A}}(\bm{r}) = \bm{A}(\bm{r}) - \langle \bm{A} (\bm{r}) \rangle_t$.
As shown above, $\langle J_R(\bm{r}) \rangle_t$ is proportional to $\bm{v}(\bm{r} ,t)$. Similarly,  the $N_\gamma$, $N$ and $E_N$ components of $\bm{A}$ are even in the momentum, and these contributions to Eq.~(\ref{eq:proj-J-R}) are also proportional to $\bm{v}(\bm{r} ,t)$. The exception is the $\bm{g}_N$ component of $\bm{A}$. Dropping contributions proportional to $\bm{v}(\bm{r},t)$ in the projected reactive flux for the reasons given above, we have
\begin{eqnarray*}
%\label{eq:proj-J-R-approx}
&&J^D_{R,t}(\bm{r})= -\sum_{i=1}^N \Theta_i^R\sum_{k=1}^{n_a} \frac{\bm{p}^\ddagger_{(k)i}}{m_k} \cdot \big( \bm{\nabla}_{r_{(k)i}}\xi_i(\bm{r}_i^{n_a}) \big)\nonumber \\
&&\qquad \qquad \times \delta(\xi_i(\bm{r}_i^{n_a})-\xi^\ddagger)\Delta(\bm{r}_{ic}-\bm{r})\nonumber \\
&&+\Big\langle \sum_{i=1}^N \Theta_i^R\sum_{k=1}^{n_a} \big( \bm{\nabla}_{r_{(k)i}}\xi_i(\bm{r}_i^{n_a}) \big)\delta(\xi_i(\bm{r}_i^{n_a})-\xi^\ddagger) \Delta(\bm{r}_{ic}-\bm{r}) \Big\rangle_t \nonumber \\
&& \qquad \qquad \cdot \frac{1}{m} \bm{g}^\ddagger_N(\bm{r}).
\end{eqnarray*}

We require a more explicit expression for
\begin{equation*}
{L}_R(\bm{r}_\parallel ,t)= \frac{1}{A}\int_0^\infty d \tau \; \langle J^D_{R,t}(\tau) J^D_{R,t}\rangle_H(\bm{r}_\parallel ,t).
\end{equation*}
We make several approximations to obtain an estimate for this transport property. The fluid velocity contributions to the dissipative fluxes will be neglected, which should be valid for the low Reynolds number conditions of interest. As discussed above the projected time evolution guarantees that the reactive flux correlation decays to zero on long time scales so that the infinite time integral is well-behaved. If there is a time scale separation between the chemical and other microscopic processes, the expression can be simplified using unprojected dynamics provided a plateau-value calculation determines the rate coefficient. In this approximation, we can write
\begin{equation*}
{L}_R(\bm{r}_\parallel ,t) \approx \frac{1}{A}\int_0^{t^*} d \tau \; \langle J^D_{R,t}(\tau) J^D_{R,t}\rangle_H(\bm{r}_\parallel ,t),
\end{equation*}
where the time integral of the reactive flux correlation will plateau at a time $t^*$ such that $ t_{\rm mic} \ll t^* \ll t_{\rm chem}$ and then decays slowly to zero as $t$ increases. The reaction rate coefficient may then be estimated from the reactive flux correlation evolving by ordinary dynamics, $J^D_{R,t}(\tau) = \exp\{i {\mathcal L}_0\tau\} J^D_{R,t}$. Making use of the fact that the reactive species are dilutely dispersed in the solution, the definition $\theta_i^A(\xi_i)= \Theta_i^R H(\xi^\ddagger-\xi_i)$, and its time derivative, and, given the short time scale of passage from the barrier top, the replacement $\bm{r}_{ic}(\tau) \approx \bm{r}_{ic}$, we can write this expression as
\begin{eqnarray*}
&&{L}_R(\bm{r}_\parallel ,t)
=- \frac{H_c(\bm{r}_\parallel)}{A} \\
&& \times\int_0^t d \tau \; \Big\langle \sum_{i=1}^N \Theta_i^R \dot{H}(\xi^\ddagger-\xi_i(\tau))\dot{\xi_i} \delta(\xi^\ddagger-\xi_i) \Big\rangle_H(\bm{r}_\parallel ,t) .\nonumber
\end{eqnarray*}
The change from reactant to product (or vice versa) will not occur unless reactive molecule $i$ is close to a catalytic site, say $\alpha$. Therefore, given that the particle is in the interaction zone, the vector distance of the parallel component of the center of a reactive molecule from the site $\alpha$, $\bm{r}_{i,\parallel}^\alpha= \bm{r}_{ic,\parallel}-R_J \hat{\bm{S}}^\alpha (\bm{R})$, is restricted to lie within the interaction zone of site $\alpha$. Then $\bm{r}_\parallel$ is restricted to the domain where the catalytic sites lie. The Heaviside function $H_c(\bm{r}_\parallel)$ accounts for this restriction. With this result in hand, we can write the reactive flux in terms of the affinity ${\cal A} = \widetilde{\mu}_B - \widetilde{\mu}_A $ as,
\begin{align*}
{\mathcal J}^{(0)}_{\gamma 0}(\bm{r}_\parallel,t) &= - \nu_\gamma  \beta {L}^R_{t}(\bm{r}_\parallel ,t) \, {\cal A} (z_0,\bm{r}_\parallel,t) \nonumber \\
&= \nu_\gamma H(\bm{r}_\parallel) \Big( \frac{{L}_R(\bm{r}_\parallel,t)}{n_A^{\rm eq}} n_A (z_0,\bm{r}_\parallel ,t) \\
&\qquad - \frac{{L}_R(\bm{r}_\parallel ,t)}{n_{B}^{\rm eq}} n_{B}(z_0,\bm{r}_\parallel ,t)\Big) \nonumber \\
&= \nu_\gamma H(\bm{r}_\parallel ) \big(\kappa_+(\bm{r}_\parallel ,t) n_A(z_0,\bm{r}_\parallel ,t) \\
& \qquad -\kappa_-(\bm{r}_\parallel ,t) n_B(z_0,\bm{r}_\parallel ,t) \big),\nonumber
\end{align*}
where the local chemical potentials $\widetilde{\mu}_\gamma$ have been approximated in the dilute limit $n_\gamma \ll n$ as
\begin{align*}
\beta \widetilde{\mu}_{\gamma} (z_0,\bm{r}_\parallel,t) &= \beta \widetilde{\mu}_{\gamma}^{\rm eq} + (n_\gamma (z_0,\bm{r}_\parallel,t) - n_\gamma^{\rm{eq}})/n_\gamma^{\rm{eq}} .
\end{align*}
The last equality defines the local rate coefficients $\kappa_{\pm}$ and provides a term in the reaction-diffusion equation that leads to the radiation boundary conditions Eq.~(\ref{eq:species-BC}) on the species density fields, 
\begin{align*}
%\label{eq:reactionRate}
{\cal R}(z_0,\bm{r}_\parallel ,t) &= H(\bm{r}_\parallel ) \bigg( \kappa_+(\bm{r}_\parallel ,t) n_A(z_0,\bm{r}_\parallel ,t) \nonumber \\
& \qquad -\kappa_-(\bm{r}_\parallel ,t) n_B(z_0,\bm{r}_\parallel ,t) \bigg).
\end{align*}

\bibliography{Janus-micro-refs}

\begin{thebibliography}{54}
\bibitem [{coa()}]{coarse}%
  \BibitemOpen
  \href@noop {} {}\bibinfo {note} {A simple example of a coarse-graining
  function can be given by considerting the Fourier transform of
  Eq.~(\ref{eq:A-Am-relation}): $A(\bm{k})=\Delta(\bm{k})A_m(\bm{k})$. If the
  local microscopic fields are restricted to small wavenumbers, $k < k_c$, so
  that they are slowly varying in space, we have $\Delta(\bm{k})=H(k_c-k)$,
  where $H(k)$ is the Heaviside function. In position space we then have
  $\Delta(\bm{r}) \sim j_1(k_c r)/(k_c r)$.}\BibitemShut {Stop}%
\end{thebibliography}

\end{document}